\documentclass[ letterpaper,english,reprint,nofootinbib,aps,superscriptaddress,showpacs]{revtex4-1}
\usepackage{babel,calc,amsmath,amsthm,amssymb,graphicx,subfigure,xcolor}
\usepackage{verbatim}
\usepackage[T1]{fontenc}
\usepackage{dsfont}
\usepackage[unicode=true]{hyperref}
\hypersetup{
     colorlinks=true,       		
     linkcolor=blue,          	
     citecolor=red,            
     urlcolor=magenta,           	
 }
\newtheorem{theorem}{Theorem}
\newtheorem*{theorem*}{Theorem}
\newtheorem{corollary}{Corollary}
\newtheorem*{corollary*}{Corollary}

\newtheorem*{lemma*}{Lemma}

\newtheorem*{proposition*}{Proposition}
\theoremstyle{definition}

\newtheorem*{definition*}{Definition}
\theoremstyle{remark}
\newtheorem{remark}{Remark}
\newtheorem*{remark*}{Remark}

\begin{document}

\title{Einstein-Podolsky-Rosen steering paradox ``2=1'' for $N$ qubits}

\author{Zhi-Jie Liu}
\affiliation{Theoretical Physics Division, Chern Institute of Mathematics and LPMC, Nankai University, Tianjin 300071, People's Republic of China}

\author{Jie Zhou}
\affiliation{College of Physics and Materials Science, Tianjin Normal University, Tianjin 300382,
   People's Republic of China}

\author{Hui-Xian Meng}
\affiliation{School of Mathematics and Physics, North China Electric Power University, Beijing 102206,
   People's Republic of China}

   \author{Xing-Yan Fan}
   \affiliation{Theoretical Physics Division, Chern Institute of Mathematics and LPMC, Nankai University, Tianjin 300071,
      People's Republic of China}

\author{Mi Xie}
\email{xiemi@tju.edu.cn}
\affiliation{Department of Physics, School of Science, Tianjin University, Tianjin 300072,  People's Republic of China
}

\author{Fu-lin Zhang}
   \email{flzhang@tju.edu.cn}
   \affiliation{Department of Physics, School of Science, Tianjin University, Tianjin 300072,  People's Republic of China
   }

\author{Jing-Ling~Chen}
\email{chenjl@nankai.edu.cn}
\affiliation{Theoretical Physics Division, Chern Institute of Mathematics and LPMC, Nankai University, Tianjin 300071,
   People's Republic of China}

\begin{abstract}
   Einstein-Podolsky-Rosen (EPR) paradox highlights the absence of a local
   realistic explanation for quantum mechanics, and shows the incompatibility of
   the local-hidden-state models with quantum theory. For $N$-qubit states, or
   more importantly, the $N$-qubit mixed states, we present the EPR steering
   paradox in the form of the contradictory equality \textquotedblleft%
   2=1\textquotedblright. We show that the contradiction holds for any $N$-qubit
   state as long as both \textquotedblleft the pure state
   requirement\textquotedblright\ and \textquotedblleft the measurement
   requirement\textquotedblright\ are satisfied. This also indicates that the EPR
   steering paradox exists in more general cases. Finally, we give specific
   examples to demonstrate and analyze our arguments.

   Keywords: Einstein-Podolsky-Rosen steering; quantum paradox;  $N$ qubits states.

\end{abstract}

\maketitle

\section{Introduction}

The quantum paradox serves as a powerful tool in elucidating the fundamental distinction between the quantum theory and the classical theory. Quantum correlations
play a central role in the study of quantum information and quantum mechanics.
Among the quantum correlations, quantum entanglement and Bell's nonlocality
are the first to be proposed and studied. In 1935, Einstein, Podolsky, and
Rosen (EPR) published their famous article \textquotedblleft\emph{Can
Quantum-Mechanical Description of Physical Reality Be Considered
Complete?}\textquotedblright\ \cite{EPR1935} , which questioned the
completeness of quantum mechanics under the assumptions of locality and
reality. This is nowadays well-known as the EPR paradox. Soon after the
publication of the EPR paper, Schr\"{o}dinger introduced two important
concepts, namely, quantum entanglement and quantum steering \cite{Schr1935,
Schr1936}. Quantum entanglement distinguishes quantum theory from classical
theory. And quantum steering is closely related to \textquotedblleft the
spooky action at a distance\textquotedblright. However, the idea of steering
did not receive considerable attention or advancement until the year 2007, at
which point Wiseman \emph{et al.} presented a meticulous definition by
utilizing quantum information concepts \cite{Wiseman2007,jones2007}. So far,
quantum steering has been widely applied in various fields
\cite{preskill1998,Galindo2002, Ekert1996,Childs2010}.

Steering is a quantum correlation between entanglement \cite{Quantum
entanglement2009} and Bell nonlocality \cite{Bell nonlocality2014, Quantum
machine learning2017,bell1964}. Steering is used to describe the situation in
a bipartite system. When people use different observables to detect one of the
particles, it will cause the corresponding other particle to collapse to a
certain state. In practice, Alice prepares a bipartite quantum state, and she
sends one of the particles to Bob. They each measure the particles in their
hands and communicate over a classical channel. It is Alice's task to convince
Bob that Alice has prepared a pair of entangled state. In the process, Bob
needs to assess the correctness of the assumptions of quantum mechanics and
acknowledge the measurements as described by quantum mechanics. Specifically,
Bob can disbelieve Alice's equipment and measurements. However, in this case,
Bob needs to rule out the influence of hidden variables on the measurement
results by the measurements he has in hand. Bob can fully trust his own
equipment and results. If Bob cannot explain the measurement results on his
side with the local-hidden states (LHS), he must recognize that Alice has
prepared an entangled bipartite state. Only EPR steering states can accomplish
this task. And quantum steering is an asymmetric quantum nonlocality. That is,
in some cases Alice can steer Bob, but in turn Bob cannot steer Alice
\cite{Asymmetric2010,Bowles2014, Quintino2015,Wollmann2016,HV2012}. Based on
some properties of quantum steering, Chen \emph{et al.} proposed the EPR
steering paradox \textquotedblleft2=1\textquotedblright\ \cite{chenjl2016},
where \textquotedblleft2\textquotedblright\ is the quantum result and
\textquotedblleft1\textquotedblright\ is the corresponding result of LHS
models. They verified the EPR steering state by the contradiction between
quantum mechanics and classical theory. In the 2-setting EPR steering
protocol, they found that any 2-qubit entangled pure state possesses the
contradiction. Thereafter, Liu \emph{et al.} found that such a contradiction
was also valid for a specific 4-qubit entangled mixed state \cite{Liu2021}. In
other words, the discussion of the EPR steering paradox \textquotedblleft%
2=1\textquotedblright\ has been limited to arbitrary 2-qubit pure state and a
special 4-qubit mixed state.

The purpose of this paper is to study the EPR steering paradox
\textquotedblleft2=1\textquotedblright\ for $N$ qubits. Based on the 2-setting
steering protocol, we have obtained such EPR steering paradox
\textquotedblleft2=1\textquotedblright\ for $N$-qubit entangled states. In
this work, we demonstrate that any $N$-qubit state can lead to the
contradiction, provided that both \textquotedblleft the pure state
requirement\textquotedblright\ and \textquotedblleft the measurement
requirement\textquotedblright\ are fulfilled simultaneously. The paper is
organized as follows. In Sec. II, we propose a theorem for $N$-qubit quantum
states that contains two requirements : \textquotedblleft the pure state
requirement\textquotedblright\ and \textquotedblleft the measurement
requirement\textquotedblright. In Sec. III, We obtain the EPR steering paradox
\textquotedblleft2=1\textquotedblright\ for the $N$-qubit states under
Bell-like basis measurement. In Sec. IV, we apply the results to the 2-qubit
mixed states and obtain a corollary that there is no EPR steering paradox
\textquotedblleft2=1\textquotedblright\ for 2-qubit mixed states. Finally, we
conclude with a summary in Sec. V. Some detailed proofs are given in the Appendix A and B.

\section{EPR steering paradox ``2=1'' for $N$-qubit states}

\subsection{``2=1'' for the 2-qubit pure state and the 4-qubit mixed state}

To make the paper be self-contained, in this subsection let us make a brief
review. In 2016, Chen \emph{et al.} first simplified the EPR steering paradox
as a contradiction \textquotedblleft2=1\textquotedblright\ for any 2-qubit
pure entangled state~\cite{chenjl2016}. They analyzed a 2-qubit pure entangled
state given by
\begin{equation}
\rho_{AB}=\left\vert \Psi\left(  \theta\right)  \right\rangle \left\langle
\Psi\left(  \theta\right)  \right\vert ,
\end{equation}
where
\begin{equation}
\left\vert \Psi\left(  \theta\right)  \right\rangle =\cos\theta\left\vert
0\right\rangle \left\vert 0\right\rangle +\sin\theta\left\vert 1\right\rangle
\left\vert 1\right\rangle , \label{eq2.01}%
\end{equation}
with $\theta\in\left(  0,\pi/2\right)  $. They chose the 2-setting protocol as
$\left\{  \hat{z},\hat{x}\right\}  $. In the protocol, Bob asks Alice to carry
out either one of two possible projective measurements on her qubit along the
$z$-direction and the $x$-direction, i.e.,
\begin{equation}%
\begin{array}
[c]{l}%
{{P}_{0}^{\hat{z}}=}\left\vert 0\right\rangle \left\langle 0\right\vert ,\\
{{P}_{1}^{\hat{z}}=}\left\vert 1\right\rangle \left\langle 1\right\vert ,\\
{{P}_{0}^{\hat{x}}=}\left\vert +\right\rangle \left\langle +\right\vert ,\\
{{P}_{1}^{\hat{x}}=}\left\vert -\right\rangle \left\langle -\right\vert ,
\end{array}
\label{eq2.02}%
\end{equation}
where $\left\vert \pm\right\rangle =(1/\sqrt{2})(  \left\vert
0\right\rangle \pm\left\vert 1\right\rangle )  $, and to inform him of
the measurement results of \textquotedblleft$a$\textquotedblright\ (where
$a=0,1$). Then Bob's four unnormalized conditional states are
\begin{equation}%
\begin{array}
[c]{l}%
\tilde{\rho}{_{0}^{\hat{z}}=\cos}^{2}\theta\left\vert 0\right\rangle
\left\langle 0\right\vert ,\quad \tilde{\rho}{_{1}^{\hat{z}}=\sin}^{2}{\theta}\left\vert 1\right\rangle
\left\langle 1\right\vert ,\\
\tilde{\rho}{_{0}^{\hat{x}}=}\dfrac{1}{2}\left(  \cos\theta\left\vert
0\right\rangle +{\sin\theta}\left\vert 1\right\rangle \right)  \left(
\cos\theta\left\langle 0\right\vert +{\sin\theta}\left\langle 1\right\vert
\right)  ,\\ [9pt]
\tilde{\rho}{_{1}^{\hat{x}}=}\dfrac{1}{2}\left(  \cos\theta\left\vert
0\right\rangle -{\sin\theta}\left\vert 1\right\rangle \right)  \left(
\cos\theta\left\langle 0\right\vert -{\sin\theta}\left\langle 1\right\vert
\right)  .
\end{array}
\label{eq2.03}%
\end{equation}

If Bob's states have a LHS description, then there exists an ensemble
$\{   \wp_{\xi}\rho_{\xi}\}  $ and a stochastic map $\wp(
a|\hat{n},\xi)  $ satisfying
\begin{equation}
\tilde{\rho}_{a}^{\hat{n}}=%
{\displaystyle\sum\limits_{\xi}}
\wp\left(  a|\hat{n},\xi\right)  \wp_{\xi}\rho_{\xi},\label{eq2.1}%
\end{equation}
and
\begin{equation}%
{\displaystyle\sum\limits_{\xi}}
\wp_{\xi}\rho_{\xi}=\rho_{B}.\label{eq2.2}%
\end{equation}
Here $\wp_{\xi}$ and $\wp\left(  a|\hat{n},\xi\right)  $ are probabilities
satisfying%
\begin{equation}
{\sum\limits_{\xi}}\wp_{\xi}=1,
\end{equation}
and$\;$%
\begin{equation}
{\sum\limits_{a}}\wp\left(  a|\hat{n},\xi\right)  =1,
\end{equation}
for a fixed $\xi$, and $\rho_{B}=\mathrm{tr}_{A}\left(  \rho_{AB}\right)  $ is
Bob's reduced density matrix \cite{Wiseman2007,jones2007}.

Then Bob's four unnormalized conditional states satisfy Eq. (\ref{eq2.1}), and one
has%
\begin{equation}%
\begin{array}
[c]{c}%
\widetilde{\rho}_{0}^{\hat{z}}=%
{\displaystyle\sum\limits_{\xi}}
\wp\left(  0|\hat{z},\xi\right)  \wp_{\xi}\rho_{\xi},\\
\widetilde{\rho}_{1}^{\hat{z}}=%
{\displaystyle\sum\limits_{\xi}}
\wp\left(  1|\hat{z},\xi\right)  \wp_{\xi}\rho_{\xi},\\
\widetilde{\rho}_{0}^{\hat{x}}=%
{\displaystyle\sum\limits_{\xi}}
\wp\left(  0|\hat{x},\xi\right)  \wp_{\xi}\rho_{\xi},\\
\widetilde{\rho}_{1}^{\hat{x}}=%
{\displaystyle\sum\limits_{\xi}}
\wp\left(  1|\hat{x},\xi\right)  \wp_{\xi}\rho_{\xi}.
\end{array}
\label{eq2.003}%
\end{equation}
Because the four states of Eq. (\ref{eq2.03}) are pure states, it is
sufficient to take $\xi$ from $1$ to $4$. Eq. (\ref{eq2.003}) can be write as
\begin{widetext}
\begin{equation}%
\begin{array}
[c]{c}%
\widetilde{\rho}_{0}^{\hat{z}}=\wp\left(  0|\hat{z},1\right)  \wp_{1}\rho
_{1}+\wp\left(  0|\hat{z},2\right)  \wp_{2}\rho_{2}+\wp\left(  0|\hat
{z},3\right)  \wp_{3}\rho_{3}+\wp\left(  0|\hat{z},4\right)  \wp_{4}\rho
_{4},\\
\widetilde{\rho}_{1}^{\hat{z}}=\wp\left(  1|\hat{z},1\right)  \wp_{1}\rho
_{1}+\wp\left(  1|\hat{z},2\right)  \wp_{2}\rho_{2}+\wp\left(  1|\hat
{z},3\right)  \wp_{3}\rho_{3}+\wp\left(  1|\hat{z},4\right)  \wp_{4}\rho
_{4},\\
\widetilde{\rho}_{0}^{\hat{x}}=\wp\left(  0|\hat{x},1\right)  \wp_{1}\rho
_{1}+\wp\left(  0|\hat{x},2\right)  \wp_{2}\rho_{2}+\wp\left(  0|\hat
{x},3\right)  \wp_{3}\rho_{3}+\wp\left(  0|\hat{x},4\right)  \wp_{4}\rho
_{4},\\
\widetilde{\rho}_{1}^{\hat{x}}=\wp\left(  1|\hat{x},1\right)  \wp_{1}\rho
_{1}+\wp\left(  1|\hat{x},2\right)  \wp_{2}\rho_{2}+\wp\left(  1|\hat
{x},3\right)  \wp_{3}\rho_{3}+\wp\left(  1|\hat{x},4\right)  \wp_{4}\rho_{4}.
\end{array}
\label{eq2.005}%
\end{equation}
\end{widetext}And owing to the fact that a pure state cannot be obtained by
convex combination of other pure states, one has
\begin{equation}%
\begin{array}
[c]{l}%
\tilde{\rho}{_{0}^{\hat{z}}=}\wp_{1}\rho_{1},\\
\tilde{\rho}{_{1}^{\hat{z}}=}\wp_{2}\rho_{2},\\
\tilde{\rho}{_{0}^{\hat{x}}=}\wp_{3}\rho_{3},\\
\tilde{\rho}{_{1}^{\hat{x}}=}\wp_{4}\rho_{4}.
\end{array}
\label{eq2.04}%
\end{equation}
Here%
\[
\wp\left(  0|\hat{z},1\right)  =\wp\left(  1|\hat{z},2\right)  =\wp\left(
0|\hat{x},3\right)  =\wp\left(  1|\hat{x},4\right)  =1,
\]
and other $\wp\left(  a|\hat{n},\xi\right)  =0$. By summing four terms up in
Eq. (\ref{eq2.04}) and taking trace, the left side gives $\mathrm{tr}(
\tilde{\rho}{^{\hat{z}}}+\tilde{\rho}{^{\hat{x}}})  =2\mathrm{tr}(
\rho_{B})  =2$. But the right side gives $\mathrm{tr}(  \wp_{1}%
\rho_{1}+\wp_{2}\rho_{2}+\wp_{3}\rho_{3}+\wp_{4}\rho_{4})
=\mathrm{tr}(  \rho_{B})  =1$, then one can obtain the
contradiction \textquotedblleft2=1\textquotedblright, i.e., the EPR steering
paradox, for any 2-qubit pure entangled state.

After that, in 2021, Liu \emph{et al.} found a 4-qubit mixed entangled state
\begin{equation}
\rho\left(  \theta\right)  =\cos^{2}\theta\left\vert LC_{4}\right\rangle
\left\langle LC_{4}\right\vert +\sin^{2}\theta\left\vert LC_{4}^{\prime
}\right\rangle \left\langle LC_{4}^{\prime}\right\vert , \label{eq2.05}%
\end{equation}
where
\begin{equation}%
\begin{array}
[c]{c}%
\left\vert LC_{4}\right\rangle =\dfrac{1}{2}\left(  \left\vert
0000\right\rangle +\left\vert 1100\right\rangle +\left\vert 0011\right\rangle
-\left\vert 1111\right\rangle \right)  ,\\ [9pt]
\left\vert LC_{4}^{\prime}\right\rangle =\dfrac{1}{2}\left(  \left\vert
0100\right\rangle +\left\vert 1000\right\rangle +\left\vert 0111\right\rangle
-\left\vert 1011\right\rangle \right)  ,
\end{array}
\label{eq2.06}%
\end{equation}
are the linear cluster states \cite{Liu2021}. Alice prepares the state
$\rho\left(  \theta\right)  $ as in Eq. (\ref{eq2.05}). She keeps 1,2
particles and sends 3,4 particles to Bob. In the 2-setting steering protocol
$\left\{  \hat{n}_{1},\hat{n}_{2}\right\}  $ $\left(  \hat{n}_{1}\neq\hat
{n}_{2}\right)  $, with%
\begin{equation}%
\begin{array}
[c]{c}%
\hat{n}_{1}=\sigma_{z}\sigma_{z}\equiv zz,\\
\hat{n}_{2}=\sigma_{y}\sigma_{x}\equiv yx.
\end{array}
\label{eq2.07}%
\end{equation}
In the protocol, Bob asks Alice to carry out either one of two possible
projective measurements on her qubits, i.e.,%
\begin{equation}%
\begin{array}
[c]{l}%
P_{00}^{\hat{n}_{1}}=\left\vert 00\right\rangle \left\langle 00\right\vert ,\\
P_{01}^{\hat{n}_{1}}=\left\vert 01\right\rangle \left\langle 01\right\vert ,\\
P_{10}^{\hat{n}_{1}}=\left\vert 10\right\rangle \left\langle 10\right\vert ,\\
P_{11}^{\hat{n}_{1}}=\left\vert 11\right\rangle \left\langle 11\right\vert ,\\
P_{00}^{\hat{n}_{2}}=\left\vert \uparrow+\right\rangle \left\langle
\uparrow+\right\vert ,\\
P_{01}^{\hat{n}_{2}}=\left\vert \uparrow-\right\rangle \left\langle
\uparrow-\right\vert ,\\
P_{10}^{\hat{n}_{2}}=\left\vert \downarrow+\right\rangle \left\langle
\downarrow+\right\vert ,\\
P_{11}^{\hat{n}_{2}}=\left\vert \downarrow-\right\rangle \left\langle
\downarrow-\right\vert ,
\end{array}
\label{eq2.08}%
\end{equation}
where $\vert \pm\rangle =(1/\sqrt{2})(\left\vert
0\right\rangle \pm\left\vert 1\right\rangle)   $, $\vert
\updownarrow\rangle =(1/\sqrt{2}) (\left\vert 0\right\rangle
\pm {\rm i}\left\vert 1\right\rangle)   $. After Alice's measurement, Bob's
unnormalized conditional states are
\begin{equation}%
\begin{array}
[c]{l}%
\tilde{\rho}_{00}^{\hat{n}_{1}}=\dfrac{1}{4}\cos^{2}\theta\left(  \left\vert
00\right\rangle +\left\vert 11\right\rangle \right)  \left(  \left\langle
00\right\vert +\left\langle 11\right\vert \right)  ,\\ [9pt]
\tilde{\rho}_{01}^{\hat{n}_{1}}=\dfrac{1}{4}\sin^{2}\theta\left(  \left\vert
00\right\rangle +\left\vert 11\right\rangle \right)  \left(  \left\langle
00\right\vert +\left\langle 11\right\vert \right)  ,\\ [9pt]
\tilde{\rho}_{10}^{\hat{n}_{1}}=\dfrac{1}{4}\sin^{2}\theta\left(  \left\vert
00\right\rangle -\left\vert 11\right\rangle \right)  \left(  \left\langle
00\right\vert -\left\langle 11\right\vert \right)  ,\\ [9pt]
\tilde{\rho}_{11}^{\hat{n}_{1}}=\dfrac{1}{4}\cos^{2}\theta\left(  \left\vert
00\right\rangle -\left\vert 11\right\rangle \right)  \left(  \left\langle
00\right\vert -\left\langle 11\right\vert \right)  ,\\ [9pt]
\tilde{\rho}_{00}^{\hat{n}_{2}}=\dfrac{1}{8}\left(  \left\vert 00\right\rangle
+{\rm i}\left\vert 11\right\rangle \right)  \left(  \left\langle 00\right\vert
-{\rm i}\left\langle 11\right\vert \right)  ,\\ [9pt]
\tilde{\rho}_{01}^{\hat{n}_{2}}=\dfrac{1}{8}\left(  \left\vert 00\right\rangle
-{\rm i}\left\vert 11\right\rangle \right)  \left(  \left\langle 00\right\vert
+{\rm i}\left\langle 11\right\vert \right)  ,\\ [9pt]
\tilde{\rho}_{10}^{\hat{n}_{2}}=\dfrac{1}{8}\left(  \left\vert 00\right\rangle
-{\rm i}\left\vert 11\right\rangle \right)  \left(  \left\langle 00\right\vert
+{\rm i}\left\langle 11\right\vert \right)  ,\\ [9pt]
\tilde{\rho}_{11}^{\hat{n}_{2}}=\dfrac{1}{8}\left(  \left\vert 00\right\rangle
+{\rm i}\left\vert 11\right\rangle \right)  \left(  \left\langle 00\right\vert
-{\rm i}\left\langle 11\right\vert \right)  .
\end{array}
\label{2.09}%
\end{equation}

If Bob's states have a LHS description, they must satisfy Eq. (\ref{eq2.1})
and Eq. (\ref{eq2.2}). Because the eight states of Eq. (\ref{2.09}) are pure
states, it is sufficient to take $\xi$ from $1$ to $8$. And a pure state
cannot be obtained by convex combination of other pure states, one has%
\begin{equation}%
\begin{array}
[c]{c}%
\tilde{\rho}_{00}^{\hat{n}_{1}}=\wp_{1}\rho_{1}  ,\\
\tilde{\rho}_{01}^{\hat{n}_{1}}=\wp_{2}\rho_{2}  ,\\
\tilde{\rho}_{10}^{\hat{n}_{1}}=\wp_{3}\rho_{3}  ,\\
\tilde{\rho}_{11}^{\hat{n}_{1}}=\wp_{4}\rho_{4}  ,\\
\tilde{\rho}_{00}^{\hat{n}_{2}}=\wp_{5}\rho_{5}  ,\\
\tilde{\rho}_{01}^{\hat{n}_{2}}=\wp_{6}\rho_{6}  ,\\
\tilde{\rho}_{10}^{\hat{n}_{2}}=\wp_{7}\rho_{7}  ,\\
\tilde{\rho}_{11}^{\hat{n}_{2}}=\wp_{8}\rho_{8}  .%
\end{array}
\label{eq2.10}%
\end{equation}
By summing eight terms up in Eq. (\ref{eq2.10}) and taking trace, the left
side gives $\mathrm{tr}(  \tilde{\rho}^{\hat{n}_{1}}+\tilde{\rho}%
^{\hat{n}_{2}})  =2\mathrm{tr}\left(  \rho_{B}\right)  =2$. But the
right side gives $\mathrm{tr}\big(
{\sum_{\xi=1}^{8}}
\wp_{\xi}\rho_{\xi}\big)  =\mathrm{tr}(  \rho_{B})  =1$, then one
can obtained an EPR steering paradox \textquotedblleft2=1\textquotedblright%
\ for the specific 4-qubit mixed entangled state.

\subsection{\textquotedblleft2=1\textquotedblright\ for $N$-qubit states}

In this subsection, we study the EPR steering paradox \textquotedblleft%
2=1\textquotedblright\ for $N$-qubit states. Our result is the following theorem.

\begin{theorem}
In the 2-setting steering protocol $\left\{  \hat{n}_{1},\hat{n}_{2}\right\}
$, Alice and Bob share an $N$-qubit state $\rho_{AB}$. Assume that Alice
measures along $\hat{n}_{1}$ and $\hat{n}_{2}$, and then Bob obtains
$\tilde{\rho}_{a}^{\hat{n}_{1}}$ and $\tilde{\rho}_{a^{\prime}}^{\hat{n}_{2}}%
$, respectively, where $a,a^{\prime}$ is the measurement result of Alice.
There will be a contradiction of \textquotedblleft2=1\textquotedblright\ if
$\rho_{AB}$ satisfies simultaneously \textquotedblleft the pure state
requirement\textquotedblright\ and \textquotedblleft the measurement
requirement\textquotedblright. The two requirements are:

\begin{enumerate}
\item The pure state requirement: Bob's unnormalized conditional states
$\{  \tilde{\rho}_{a}^{\hat{n}_{1}}\}  $ and $\{  \tilde{\rho
}_{a^{\prime}}^{\hat{n}_{2}}\}  $ are all pure states.

\item The measurement requirement: any one of $\{  \tilde{\rho}_{a}%
^{\hat{n}_{1}}\}  $ is different from any one of $\{  \tilde{\rho
}_{a^{\prime}}^{\hat{n}_{2}}\}  $.
\end{enumerate}
\end{theorem}

Let Alice and Bob share an $N$-qubit entangled state
\begin{equation}
\rho_{AB}=\sum_{\alpha}p_{\alpha}\left\vert \psi_{AB}^{\left(  \alpha\right)
}\right\rangle \left\langle \psi_{AB}^{\left(  \alpha\right)  }\right\vert .
\end{equation}
Alice has $M\left(  M<N\right)  $ particles and Bob has $(N-M)$ particles. In
the 2-setting steering protocol $\left\{  \hat{n}_{1},\hat{n}_{2}\right\}  $
$\left(  \text{with }\hat{n}_{1}\neq\hat{n}_{2}\right)  $, Alice performs
$2^{M+1}$ projective measurements, each of them measuring $M$ particles of
Alice.
For each projective measurement $\hat{P}_{a}^{\hat{n}_{k}}$, Bob obtains the
corresponding unnormalized state $\tilde{\rho}_{a}^{\hat{n}_{k}}%
=\mathrm{tr}_{A}\left[ \left(  {{P}_{a}^{\hat{n}_{k}}\otimes\mathds{1}}%
\right)  \rho_{AB}\right]  $, with $\hat{n}_{k}$ the measurement direction,
$a$ the Alice's measurement result, ${\mathds{1}}$ the $2^{N-M}\times2^{N-M}$
identity matrix. The wave-function $\vert \psi_{AB}^{\left(  \alpha\right)  }\rangle $ may be written as
\begin{equation}
\left\vert \psi_{AB}^{\left(  \alpha\right)  }\right\rangle =\sum_{i}\left(
s_{i+}^{\left(  \alpha\right)  }\left\vert +\phi_{i}\right\rangle \left\vert
\eta_{i+}^{\left(  \alpha\right)  }\right\rangle +s_{i-}^{\left(
\alpha\right)  }\left\vert -\phi_{i}\right\rangle \left\vert \eta
_{i-}^{\left(  \alpha\right)  }\right\rangle \right)  ,\label{2.1}%
\end{equation}
or
\begin{equation}
\left\vert \psi_{AB}^{\left(  \alpha\right)  }\right\rangle =\sum_{j}\left(
t_{j+}^{\left(  \alpha\right)  }\left\vert +\varphi_{j}\right\rangle
\left\vert \varepsilon_{j+}^{\left(  \alpha\right)  }\right\rangle
+t_{j-}^{\left(  \alpha\right)  }\left\vert -\varphi_{j}\right\rangle
\left\vert \varepsilon_{j-}^{\left(  \alpha\right)  }\right\rangle \right)
,\label{2.2}%
\end{equation}
where $i,j=1,2,\cdots,2^{M-1}$. Eq. (\ref{2.1}) and Eq. (\ref{2.2}) are two
representations of $\left\vert \psi_{AB}^{  \alpha
}\right\rangle $ in different bases. Here $\left\vert \pm\phi_{i}\right\rangle
$ and $\left\vert \pm\varphi_{j}\right\rangle $\ are the eigenstates of the
operator $\hat{P}_{a}^{\hat{n}_{k}}$ $\left(  \text{with }k=1,2\right)  $,
respectively. And $\left\{  \left\vert \pm\phi_{i}\right\rangle \right\}  $
and $\left\{  \left\vert \pm\varphi_{j}\right\rangle \right\}  $ are two sets
of complete basis of $2^{M}$-dimensional Hilbert space, respectively. At the
same time, $\left\vert \pm\phi_{i}\right\rangle $ and $\left\vert \pm
\varphi_{j}\right\rangle $ also represent Alice's particles. $\vert
\eta_{i\pm}^{(  \alpha)  }\rangle $ and $\vert
\varepsilon_{j\pm}^{(  \alpha)  }\rangle $ are Bob's
collapsed states (unnormalized), where

\begin{equation}
   \left\vert \eta_{i\pm}^{\left(  \alpha\right)  }\right\rangle \left\langle
\eta_{i\pm}^{\left(  \alpha\right)  }\right\vert =\text{tr}_{A}\bigg[  \left(
\left\vert \pm\phi_{i}\right\rangle \left\langle \pm\phi_{i}\right\vert
\otimes{\mathds{1}}\right)  \left\vert \psi_{AB}^{\left(  \alpha\right)
}\right\rangle \left\langle \psi_{AB}^{\left(  \alpha\right)  }\right\vert
\bigg]  
 ,
   \end{equation}
   and
   \begin{equation}
      \left\vert \varepsilon_{j\pm}^{\left(  \alpha\right)  }\right\rangle
\left\langle \varepsilon_{j\pm}^{\left(  \alpha\right)  }\right\vert
=\text{tr}_{A}\bigg[  \left(  \left\vert \pm\varphi_{j}\right\rangle
\left\langle \pm\varphi_{j}\right\vert \otimes{\mathds{1}}\right)  \left\vert
\psi_{AB}^{\left(  \alpha\right)  }\right\rangle \left\langle \psi
_{AB}^{\left(  \alpha\right)  }\right\vert \bigg] ,
   \end{equation}
where ${\mathds{1}}$ is a $2^{(  N-M)  }\times2^{(  N-M)  }$
 identity matrix.
$s_{i\pm}^{(  \alpha)  }$ and $t_{j\pm}^{(  \alpha)  }$
are complex numbers satisfying
\begin{equation}
\sum_{i}\left(  \left\vert s_{i+}^{\left(  \alpha\right)  }\right\vert
^{2}+\left\vert s_{i-}^{\left(  \alpha\right)  }\right\vert ^{2}\right)  =1,
\end{equation}
and%
\begin{equation}
\sum_{j}\left(  \left\vert t_{j+}^{\left(  \alpha\right)  }\right\vert
^{2}+\left\vert t_{j-}^{\left(  \alpha\right)  }\right\vert ^{2}\right)  =1.
\end{equation}

\emph{The two requirements can be rewritten as:}

\begin{enumerate}
\item \emph{The pure state requirement: }$\vert \eta_{i\pm}^{(
\alpha)  }\rangle $\emph{ and }$\vert \varepsilon_{j\pm
}^{(  \alpha)  }\rangle $\emph{ are independent of }$\alpha
$\emph{.}

\item \emph{The measurement requirement: for the result obtained by Bob, any
one of }$\{  \vert \eta_{i\pm}\rangle \}  $\emph{ is
different from any one of }$\{  \vert \varepsilon_{j\pm
}\rangle \}  $
\end{enumerate}

The pure state requirement guarantees that Bob's unnormalized conditional
states are all pure states. After Alice's measurement, Bob obtains the states
\begin{equation}
\tilde{\rho}_{a_{i\pm}}^{\hat{n}_{1}}=\sum_{\alpha}p_{\alpha}\left\vert
s_{i\pm}^{\left(  \alpha\right)  }\right\vert ^{2}\left\vert \eta_{i\pm
}^{\left(  \alpha\right)  }\right\rangle \left\langle \eta_{i\pm}^{\left(
\alpha\right)  }\right\vert ,
\end{equation}
and
\begin{equation}
\tilde{\rho}_{a_{j\pm}^{\prime}}^{\hat{n}_{2}}=\sum_{\alpha}p_{\alpha
}\left\vert t_{j\pm}^{\left(  \alpha\right)  }\right\vert ^{2}\left\vert
\varepsilon_{j\pm}^{\left(  \alpha\right)  }\right\rangle \left\langle
\varepsilon_{j\pm}^{\left(  \alpha\right)  }\right\vert .
\end{equation}
Since $p_{\alpha}$, $\vert s_{i\pm}^{(  \alpha)  }\vert
^{2}$ and $\vert t_{j\pm}^{(  \alpha)  }\vert ^{2}$ are
all non-negative, any terms cannot cancell. Bob's un-normalized conditional
states are pure if and only if
\begin{equation}
\left\{
\begin{array}
[c]{l}%
s_{i\pm}^{\left(  \alpha\right)  }\left\vert \eta_{i\pm}^{\left(
\alpha\right)  }\right\rangle =c_{\alpha^{\prime}\left(  i\pm\right)
}^{\alpha}s_{i\pm}^{(  \alpha^{\prime})  }\left\vert \eta_{i\pm
}^{(  \alpha^{\prime})  }\right\rangle ,\\ [9pt]
t_{j\pm}^{\left(  \alpha\right)  }\left\vert \varepsilon_{j\pm}^{\left(
\alpha\right)  }\right\rangle =d_{\alpha^{\prime}\left(  j\pm\right)
}^{\alpha}t_{j\pm}^{(  \alpha^{\prime})  }\left\vert \varepsilon
_{j\pm}^{(  \alpha^{\prime})  }\right\rangle ,
\end{array}
\right.  \label{2.4}%
\end{equation}
where $c_{\alpha^{\prime}\left(  i\pm\right)  }^{\alpha}$ and $d_{\alpha
^{\prime}\left(  j\pm\right)  }^{\alpha}$ are arbitrary complex numbers
related to $i\pm$ and $j\pm$, respectively. For convenience, we analyze in the
representation with $\left\{  \left\vert \pm\phi_{i}\right\rangle
\otimes\left\vert \eta_{i\pm}\right\rangle \right\}  $ as basis. The
projection of $\vert \psi_{AB}^{(  \alpha)  }\rangle $
in Eq. (\ref{2.1}) onto $\vert \pm\varphi_{j}\rangle $ is
\begin{align}
   \left\langle \pm\varphi_{j^{\prime}}| \psi_{AB}^{\left(  \alpha\right)
   }\right\rangle  &  =\sum_{i}s_{i\pm}^{\left(  \alpha\right)  }\left\langle
   \pm\varphi_{j^{\prime}}\left\vert \pm\phi_{i}\right.  \right\rangle \left\vert
   \eta_{i\pm}^{\left(  \alpha\right)  }\right\rangle ,\nonumber\\
   &  \equiv\sum_{i}V_{\left(  j^{\prime}\pm\right)  \left(  i\pm\right)
   }s_{i\pm}^{\left(  \alpha\right)  }\left\vert \eta_{i\pm}^{\left(
   \alpha\right)  }\right\rangle ,
   \end{align}
where $V_{(  j^{\prime}\pm)  (  i\pm)  }\equiv
\langle \pm\varphi_{j^{\prime}}  \vert \pm\phi_{i}%
\rangle $, $\{  V_{(  j^{\prime}\pm)  (
i\pm)  }\}  $ can be written as\begin{widetext}
\[
V=\left(
\begin{array}
[c]{cccccc}%
V_{\left(  1+\right)  \left(  1+\right)  } & \cdots & V_{\left(  1+\right)
\left(  2^{M-1}+\right)  } & V_{\left(  1+\right)  \left(  1-\right)  } &
\cdots & V_{\left(  1+\right)  \left(  2^{M-1}-\right)  }\\
\vdots &  & \vdots & \vdots &  & \vdots\\
V_{\left(  2^{M-1}+\right)  \left(  1+\right)  } & \cdots & V_{\left(
2^{M-1}+\right)  \left(  2^{M-1}+\right)  } & V_{\left(  2^{M-1}+\right)
\left(  1-\right)  } & \cdots & V_{\left(  2^{M-1}+\right)  \left(
2^{M-1}-\right)  }\\
V_{\left(  1-\right)  \left(  1+\right)  } & \cdots & V_{\left(  1-\right)
\left(  2^{M-1}+\right)  } & V_{\left(  1-\right)  \left(  1-\right)  } &
\cdots & V_{\left(  1-\right)  \left(  2^{M-1}-\right)  }\\
\vdots &  & \vdots & \vdots &  & \vdots\\
V_{\left(  2^{M-1}-\right)  \left(  1+\right)  } & \cdots & V_{\left(
2^{M-1}-\right)  \left(  2^{M-1}+\right)  } & V_{\left(  2^{M-1}-\right)
\left(  1-\right)  } & \cdots & V_{\left(  2^{M-1}-\right)  \left(
2^{M-1}-\right)  }%
\end{array}
\right)  .
\]
\end{widetext}
Then we have%
\begin{equation}
t_{j^{\prime}\pm}^{\left(  \alpha\right)  }\left\vert \varepsilon_{j^{\prime
}\pm}^{\left(  \alpha\right)  }\right\rangle =\sum_{i}V_{\left(  j^{\prime}%
\pm\right)  \left(  i\pm\right)  }s_{i\pm}^{\left(  \alpha\right)  }\left\vert
\eta_{i\pm}^{\left(  \alpha\right)  }\right\rangle .
\end{equation}
The pure state requirement Eq. (\ref{2.4}) can be expressed as
\begin{equation}
\left\{
\begin{array}
[c]{l}%
s_{i\pm}^{\left(  \alpha\right)  }\left\vert \eta_{i\pm}^{\left(
\alpha\right)  }\right\rangle =c_{\alpha^{\prime}\left(  i\pm\right)
}^{\alpha}s_{i\pm}^{(  \alpha^{\prime})  }\left\vert \eta_{i\pm
}^{(  \alpha^{\prime})  }\right\rangle ,\\
\left\vert \chi_{j\pm}^{\left(  \alpha\right)  }\right\rangle =d_{\alpha
^{\prime}\left(  j\pm\right)  }^{\alpha}\left\vert \chi_{j\pm}^{\left(
\alpha^{\prime}\right)  }\right\rangle .
\end{array}
\right.  \label{2.5}%
\end{equation}
Where $\vert \chi_{j\pm}^{(  \alpha)  }\rangle
\equiv\sum_{i}V_{(  j\pm)  (  i\pm)  }s_{i\pm}^{(
\alpha)  }\vert \eta_{i\pm}^{(  \alpha)  }\rangle
$.

The measurement requirement suggests that if Alice chooses different
measurements $\hat{P}_{a}^{\hat{n}_{1}}$ or $\hat{P}_{a^{\prime}}^{\hat{n}%
_{2}}$, Bob cannot get the same result.\ We prove that in Appendix A and
Appendix B the results obtained by Bob cannot be the same in different
measurements. And \textquotedblleft the different
measurements\textquotedblright\ refers to the measurements in different
directions ($\hat{n}_{1}$ and $\hat{n}_{2}$). After Alice's measurements, Bob
obtains $s_{i\pm}^{(  \alpha)  }\vert \eta_{i\pm}^{(
\alpha)  }\rangle $ and $\vert \chi_{j\pm}^{(
\alpha)  }\rangle=\sum_{i}V_{(  j\pm)  (
i\pm)  }s_{i\pm}^{(  \alpha)  }\vert \eta_{i\pm
}^{(  \alpha)  }\rangle  $. It can be seen that
$\vert \chi_{j\pm}^{(  \alpha)  }\rangle $ is obtained
by superposition of $\vert \eta_{i\pm}^{(  \alpha)
}\rangle $.\ If Bob's two results are required to be different,
$\vert \chi_{j\pm}^{(  \alpha)  }\rangle $ contains at
least two summation terms. This also requires: (1) At least two terms in the
summation in $\vert \psi_{AB}^{(  \alpha)  }\rangle $
are nonzero. That is, at least two $s_{i\pm}^{(  \alpha)  }$ in
$\vert \psi_{AB}^{(  \alpha)  }\rangle $ are nonzero.
(2) The matrix $\{  V_{(  j\pm)  (  i\pm)
}\}  $ has at least two nonzero matrix elements in each row. That is,
the two measurements ${{P}_{a}^{\hat{n}_{1}}}$ and ${{P}_{a^{\prime}}^{\hat
{n}_{2}}}$ are different. (3) $\vert \psi_{AB}^{(\alpha)
}\rangle \neq\big[  \sum_{i}(  s_{i+}^{(  \alpha)
}\vert +\phi_{i}\rangle +s_{i-}^{(  \alpha)  }\vert
-\phi_{i}\rangle )  \big]  \otimes\vert \eta_{\ell
}^{(  \alpha)  }\rangle $, where $\vert \eta_{\ell}^{(
\alpha)  }\rangle $ is one of $\big\{  \vert \eta_{i\pm
}^{(  \alpha)  }\rangle\big\}  $. That is, each
$\vert \psi_{AB}^{(  \alpha)  }\rangle $ is an entangled state.

\begin{proof}
Here, we prove that for $N$-qubit state $\rho_{AB}$, the difference between
quantum theory and classical theory can be expressed as \textquotedblleft%
2=1\textquotedblright, as long as the pure state requirement\ and the
measurement requirement\ are satisfied. It is well known that if $\rho_{AB}$
satisfies two requirements at the same time, Bob's unnormalized conditional
states are all pure states. And for different measurements $\hat{P}_{a}%
^{\hat{n}_{1}}$ and $\hat{P}_{a^{\prime}}^{\hat{n}_{2}}$, Bob 's results are
different. Without loss of generality, we assume that Bob's $2^{M+1}$
unnormalized conditional states are different. Then for the quantum results we
have%
\begin{equation}
\left\{
\begin{array}
[c]{l}%
\tilde{\rho}_{a_{1+}}^{\hat{n}_{1}}=\sum_{\alpha}p_{\alpha}\left\vert
s_{1+}^{\left(  \alpha\right)  }\right\vert ^{2}\left\vert \eta_{1+}%
\right\rangle \left\langle \eta_{1+}\right\vert ,\\
\cdots\\
\tilde{\rho}_{a_{2^{M-1}+}}^{\hat{n}_{1}}=\sum_{\alpha}p_{\alpha}\left\vert
s_{2^{M-1}+}^{\left(  \alpha\right)  }\right\vert ^{2}\left\vert \eta
_{2^{M-1}+}\right\rangle \left\langle \eta_{2^{M-1}+}\right\vert ,\\
\tilde{\rho}_{a_{2^{M-1}-}}^{\hat{n}_{1}}=\sum_{\alpha}p_{\alpha}\left\vert
s_{2^{M-1}-}^{\left(  \alpha\right)  }\right\vert ^{2}\left\vert \eta
_{2^{M-1}-}\right\rangle \left\langle \eta_{2^{M-1}-}\right\vert ,\\
\cdots\\
\tilde{\rho}_{a_{1-}}^{\hat{n}_{1}}=\sum_{\alpha}p_{\alpha}\left\vert
s_{1-}^{\left(  \alpha\right)  }\right\vert ^{2}\left\vert \eta_{1-}%
\right\rangle \left\langle \eta_{1-}\right\vert ,\\
\tilde{\rho}_{a_{1+}^{\prime}}^{\hat{n}_{2}}=\sum_{\alpha}p_{\alpha}\left\vert
t_{1+}^{\left(  \alpha\right)  }\right\vert ^{2}\left\vert \varepsilon
_{1+}\right\rangle \left\langle \varepsilon_{1+}\right\vert ,\\
\cdots\\
\tilde{\rho}_{a_{2^{M-1}+}^{\prime}}^{\hat{n}_{2}}=\sum_{\alpha}p_{\alpha
}\left\vert t_{2^{M-1}+}^{\left(  \alpha\right)  }\right\vert ^{2}\left\vert
\varepsilon_{2^{M-1}+}\right\rangle \left\langle \varepsilon_{2^{M-1}%
+}\right\vert ,\\
\tilde{\rho}_{a_{2^{M-1}-}^{\prime}}^{\hat{n}_{2}}=\sum_{\alpha}p_{\alpha
}\left\vert t_{2^{M-1}-}^{\left(  \alpha\right)  }\right\vert ^{2}\left\vert
\varepsilon_{2^{M-1}-}\right\rangle \left\langle \varepsilon_{2^{M-1}%
-}\right\vert ,\\
\cdots\\
\tilde{\rho}_{a_{1-}^{\prime}}^{\hat{n}_{2}}=\sum_{\alpha}p_{\alpha}\left\vert
t_{1-}^{\left(  \alpha\right)  }\right\vert ^{2}\left\vert \varepsilon
_{1-}\right\rangle \left\langle \varepsilon_{1-}\right\vert .
\end{array}
\right.  \label{eq2.004}%
\end{equation}

Suppose Bob's states have a LHS description, they must satisfy Eq.
(\ref{eq2.1}) and Eq. (\ref{eq2.2}). Then, Bob will check the following set of
$2^{M+1}$ equations:%
\begin{equation}
\tilde{\rho}_{a}^{\hat{n}_{k}}=%
{\displaystyle\sum\limits_{\xi}}
\wp\left(  a|\hat{n}_{k},\xi\right)  \wp_{\xi}\rho_{\xi}. \label{eq2.3}%
\end{equation}
If these $2^{M+1}$ equations have a contradiction, that is they cannot have a
common solution for the sets $\{  \wp_{\xi}\rho_{\xi}\}  $ and $\wp(
a|\hat{n}_{k},\xi)  $, then Bob is convinced that the LHS models are
non-existent and that Alice can steer the state of his qubits.

In the quantum result, there are $2^{M+1}$ pure states in Eq. (\ref{eq2.004}).
Then in the LHS description, it is sufficient to take $\xi$ from 1 to
$2^{M+1}$. It is a fact that a density matrix of pure state can only be
expanded by itself. Therefore, any $\tilde{\rho}_{a}^{\hat{n}_{k}}$ in the
equation (\ref{eq2.3}) contains only one term. So for the LHS models results
we have
\begin{equation}
\left\{
\begin{array}
[c]{l}%
\tilde{\rho}_{a_{1+}}^{\hat{n}_{1}}=\wp_{1}\rho_{1},\\
\cdots\\
\tilde{\rho}_{a_{2^{M-1}+}}^{\hat{n}_{1}}=\wp_{2^{M-1}}\rho_{2^{M-1}},\\
\tilde{\rho}_{a_{2^{M-1}-}}^{\hat{n}_{1}}=\wp_{2^{M-1}+1}\rho_{2^{M-1}+1},\\
\cdots\\
\tilde{\rho}_{a_{1-}}^{\hat{n}_{1}}=\wp_{2^{M}}\rho_{2^{M}},\\
\tilde{\rho}_{a_{1+}^{\prime}}^{\hat{n}_{2}}=\wp_{2^{M}+1}\rho_{2^{M}+1},\\
\cdots\\
\tilde{\rho}_{a_{2^{M-1}+}^{\prime}}^{\hat{n}_{2}}=\wp_{2^{M}+2^{M-1}}%
\rho_{2^{M}+2^{M-1}},\\
\tilde{\rho}_{a_{2^{M-1}-}^{\prime}}^{\hat{n}_{2}}=\wp_{2^{M}+2^{M-1}+1}%
\rho_{2^{M}+2^{M-1}+1},\\
\cdots\\
\tilde{\rho}_{a_{1-}^{\prime}}^{\hat{n}_{2}}=\wp_{2^{M+1}}\rho_{2^{M+1}}.
\end{array}
\right.  \label{eq2.4}%
\end{equation}
Finally, we sum up terms in Eq. (\ref{eq2.4}) and take the trace. The left
side gives $\mathrm{tr}(  \tilde{\rho}^{\hat{n}_{1}}+\tilde{\rho}%
^{\hat{n}_{2}})  =2\mathrm{tr}(  \rho_{B})  =2$, the result
of quantum. While the right side gives $\mathrm{tr}(  \wp_{1}\rho
_{1}+\cdots+\wp_{2^{M+1}}\rho_{2^{M+1}})  =\mathrm{tr}(  \rho
_{B})  =1$, the result of the classical LHS models. This leads to the
contradiction \textquotedblleft2=1\textquotedblright,\ which represents the
EPR paradox in the 2-setting steering protocol.
\end{proof}

\begin{remark}
For 2-qubit pure state Eq. (\ref{eq2.01}), Bob's unnormalized conditional
states are always pure. So only need to verify whether it satisfies the
measurement requirement. It can be seen that in the 2-setting protocol
$\{  \hat{z},\hat{x}\}  $, Bob 's results are $\{  \vert
0\rangle \langle 0\vert ,\vert 1\rangle
\langle 1\vert \}  $ and $\{  \vert +\rangle
\langle +\vert ,\vert -\rangle \langle -\vert
\}  $, respectively \cite{chenjl2016}. Obviously, this satisfies the
measurement requirement. According to our analysis, such state Eq.
(\ref{eq2.01}) can get the contradiction \textquotedblleft%
2=1\textquotedblright.
\end{remark}

\begin{remark}
For the 4-qubit mixed state Eq. (\ref{eq2.05}), it is necessary to analyze
whether it satisfies both the pure state requirement and the measurement
requirement. In the 2-setting protocol $\{  \hat{n}_{1},\hat{n}%
_{2}\}  =\{  \hat{z}\hat{z},\hat{y}\hat{x}\}  $, Bob's eight
conditional states are all pure states as shown in Eq. (\ref{2.09}). And for
two different measurements $P_{a}^{\hat{n}_{1}}$ and $P_{a^{\prime}}^{\hat
{n}_{2}}$, Bob's results are different. It is obvious that the pure state
requirement and the measurement requirement are satisfied at the same time, so
the specific 4-qubit mixed state Eq. (\ref{eq2.05}) can also get the
contradiction \textquotedblleft2=1\textquotedblright.
\end{remark}

In the EPR steering paradox, we propose a theorem for $N$-qubit quantum states
which contains two requirements: the pure state requirement and the
measurement requirement. If Alice and Bob share an $N$-qubit mixed state,
there will be a contradiction of \textquotedblleft2=1\textquotedblright\ only
when the pure state requirement and the measurement requirement are satisfied
at the same time. If they share an $N$-qubit pure state, the pure state
requirement is automatically satisfied. In this situation, after Alice's
measurement, Bob must get a pure state, which only needs to meet the
measurement requirement. And our results are completely consistent with the
previous conclusions. This confirms the correctness of our conclusion.

\section{Bell-like basis measurement}

Here we show a more specific example of the Bell-like basis measurement\ for
the $N$-qubit mixed states. Let us consider Alice and Bob share an $N$-qubit
entangled state $\rho_{AB}=\sum_{\alpha}p_{\alpha}\vert \psi
_{AB}^{(  \alpha)  }\rangle \langle \psi_{AB}^{(
\alpha)  }\vert $, in which$\vert \psi_{AB}^{(
\alpha)  }\rangle $ may be written as
\begin{equation}
\left\vert \psi_{AB}^{\left(  \alpha\right)  }\right\rangle =\sum_{i}\left(
s_{i+}^{\left(  \alpha\right)  }\left\vert +\phi_{i}\right\rangle \left\vert
\eta_{i+}^{\left(  \alpha\right)  }\right\rangle +s_{i-}^{\left(
\alpha\right)  }\left\vert -\phi_{i}\right\rangle \left\vert \eta
_{i-}^{\left(  \alpha\right)  }\right\rangle \right)  . \label{eq3.2}%
\end{equation}
and $\sum_{i} ( \vert s_{i+}^{(  \alpha)  }\vert
^{2}+\vert s_{i-}^{(  \alpha)  }\vert ^{2} ) =1$.
Alice has $M(  M<N)  $ particles and Bob has $(N-M)$ particles. In
particular, in the 2-setting steering protocol $\left\{  \hat{n}_{1},\hat
{n}_{2}\right\}  $ $\left(  \text{with }\hat{n}_{1}\neq\hat{n}_{2}\right)  $,
Alice performs the Bell-like basis measurement on her qubits. Then according
to the theorem, we analyze whether this example can obtain the contradiction
\textquotedblleft2=1\textquotedblright, and if yes, what conditions $\rho
_{AB}$ need to meet with the Bell-like basis measurement.

The Bell-like basis measurement can be written as\begin{widetext}
   \begin{equation}
      \left\{
      \begin{array}
      [c]{l}%
      P_{a_{1+}}^{\hat{n}_{k}}=\left(  \cos\beta_{k}\left\vert +\phi_{1}%
      \right\rangle +\sin\beta_{k}\left\vert -\phi_{1}\right\rangle \right)  \left(
      \cos\beta_{k}\left\langle +\phi_{1}\right\vert +\sin\beta_{k}\left\langle
      -\phi_{1}\right\vert \right)  ,\\
      \vdots\\
      P_{a_{2^{M-1}+}}^{\hat{n}_{k}}=\left(  \cos\beta_{k}\left\vert +\phi_{2^{M-1}%
      }\right\rangle +\sin\beta_{k}\left\vert -\phi_{2^{M-1}}\right\rangle \right)
      \left(  \cos\beta_{k}\left\langle +\phi_{2^{M-1}}\right\vert +\sin\beta
      _{k}\left\langle -\phi_{2^{M-1}}\right\vert \right)  ,\\
      P_{a_{2^{M-1}-}}^{\hat{n}_{k}}=\left(  \sin\beta_{k}\left\vert +\phi_{2^{M-1}%
      }\right\rangle -\cos\beta_{k}\left\vert -\phi_{2^{M-1}}\right\rangle \right)
      \left(  \sin\beta_{k}\left\langle +\phi_{2^{M-1}}\right\vert -\cos\beta
      _{k}\left\langle -\phi_{2^{M-1}}\right\vert \right)  ,\\
      \vdots\\
      P_{a_{1-}}^{\hat{n}_{k}}=\left(  \sin\beta_{k}\left\vert +\phi_{1}%
      \right\rangle -\cos\beta_{k}\left\vert -\phi_{1}\right\rangle \right)  \left(
      \sin\beta_{k}\left\langle +\phi_{1}\right\vert -\cos\beta_{k}\left\langle
      -\phi_{1}\right\vert \right)  ,
      \end{array}
      \right.  \label{eq3.1}%
      \end{equation}
\end{widetext}in which $\beta_{k}\in\left[  0,2\pi\right)  $. Alice performs
the measurement along $\hat{n}_{k}$ directions (with $k=1,2$). And $\vert
\pm\phi_{i}\rangle $ ( $i=1,2,\cdots,2^{M-1}$) is a set of complete
basis of $2^{M}$-dimensional Hilbert space. We prove that in Appendix C. After
Alice's measurement, Bob obtains
\begin{equation}
\left\{
\begin{array}
[c]{l}%
\tilde{\rho}_{a_{i+}}^{\hat{n}_{k}}=\sum_{\alpha}p_{\alpha}\left\vert
\omega_{i+}^{k\left(  \alpha\right)  }\right\rangle \left\langle \omega
_{i+}^{k\left(  \alpha\right)  }\right\vert ,\\ [6pt]
\tilde{\rho}_{a_{i-}}^{\hat{n}_{k}}=\sum_{\alpha}p_{\alpha}\left\vert
\omega_{i-}^{k\left(  \alpha\right)  }\right\rangle \left\langle \omega
_{i-}^{k\left(  \alpha\right)  }\right\vert ,
\end{array}
\right.  \label{eq3.3}%
\end{equation}
with
\begin{equation}%
\begin{array}
[c]{c}%
\left\vert \omega_{i+}^{k\left(  \alpha\right)  }\right\rangle =s_{i+}%
^{\left(  \alpha\right)  }\cos\beta_{k}\left\vert \eta_{i+}^{\left(
\alpha\right)  }\right\rangle +s_{i-}^{\left(  \alpha\right)  }\sin\beta
_{k}\left\vert \eta_{i-}^{\left(  \alpha\right)  }\right\rangle ,\\ [6pt]
\left\vert \omega_{i-}^{k\left(  \alpha\right)  }\right\rangle =s_{i+}%
^{\left(  \alpha\right)  }\sin\beta_{k}\left\vert \eta_{i+}^{\left(
\alpha\right)  }\right\rangle -s_{i-}^{\left(  \alpha\right)  }\cos\beta
_{k}\left\vert \eta_{i-}^{\left(  \alpha\right)  }\right\rangle .
\end{array}
\end{equation}

Firstly, in Bell-like basis measurement, the transformation matrix $\{
V_{(  j\pm)  (  i\pm)}  \}  $ is\begin{widetext}
\begin{equation}
V=\left(
\begin{array}
[c]{cccccc}%
\cos\left(  \beta_{1}-\beta_{2}\right)  & 0 & \cdots & \sin\left(  \beta
_{1}-\beta_{2}\right)  & 0 & \cdots\\
0 & \cos\left(  \beta_{1}-\beta_{2}\right)  & \cdots & 0 & \sin\left(
\beta_{1}-\beta_{2}\right)  & \cdots\\
\vdots & \vdots & \ddots & \vdots & \vdots & \vdots\\
-\sin\left(  \beta_{1}-\beta_{2}\right)  & 0 & \cdots & \cos\left(  \beta
_{1}-\beta_{2}\right)  & 0 & \cdots\\
0 & -\sin\left(  \beta_{1}-\beta_{2}\right)  & \cdots & 0 & \cos\left(
\beta_{1}-\beta_{2}\right)  & \cdots\\
\vdots & \vdots & \vdots & \vdots & \vdots & \vdots
\end{array}
\right)  . \label{eq3.4}%
\end{equation}
\end{widetext}It is obvious that there are only two nonzero matrix elements
$V_{jq+}$ as well as $V_{jq-}$ (with $q=1,2,\cdots,2^{M-1}$) in each row of matrix
$\{  V_{(  j\pm) (  i\pm)  }\}  $, and
$\vert \chi_{j\pm}^{(  \alpha)  }\rangle $ contains only
two terms. Similarly, only $\vert \eta_{q+}^{(  \alpha)
}\rangle $ and $\vert \eta_{q-}^{(  \alpha)
}\rangle $ contribute to $\vert \psi_{AB}^{(  \alpha)
}\rangle $, and we can consider $\vert \psi_{AB}^{(
\alpha)  }\rangle $ as
\begin{equation}
\left\vert \psi_{AB}^{\left(  \alpha\right)  }\right\rangle =s_{q+}^{\left(
\alpha\right)  }\left\vert +\phi_{q}\right\rangle \left\vert \eta
_{q+}^{\left(  \alpha\right)  }\right\rangle +s_{q-}^{\left(  \alpha\right)
}\left\vert -\phi_{q}\right\rangle \left\vert \eta_{q-}^{\left(
\alpha\right)  }\right\rangle . \label{eq3.5}%
\end{equation}
The measurement requirement also requires $\vert \psi_{AB}^{(
\alpha)  }\rangle \neq\big[  \sum_{i}(  s_{i+}^{(
\alpha)  }\vert +\phi_{i}\rangle +s_{i-}^{(
\alpha)  }\vert -\phi_{i}\rangle )  \big]
\otimes\vert \eta_{\ell}^{(  \alpha)  }\rangle $, so
$\vert \eta_{q+}^{(  \alpha)  }\rangle \neq\vert
\eta_{q-}^{(  \alpha)  }\rangle $. In this way, the
measurement requirement is satisfied. Secondly, the pure state requirement
requires that $\vert \eta_{q\pm}^{(  \alpha)  }\rangle $
is independent of $\alpha$, i.e.
\begin{equation}
\left\vert \psi_{AB}^{\left(  \alpha\right)  }\right\rangle =s_{q+}^{\left(
\alpha\right)  }\left\vert +\phi_{q}\right\rangle \left\vert \eta
_{q+}\right\rangle +s_{q-}^{\left(  \alpha\right)  }\left\vert -\phi
_{q}\right\rangle \left\vert \eta_{q-}\right\rangle . \label{eq3.6}%
\end{equation}

It is apparent that after a series of analysis, the form of $\vert
\psi_{AB}^{(  \alpha)  }\rangle $ is simple and only contains
two terms. There is an interesting question worthy of our further analysis,
that is, whether $\vert \psi_{AB}^{(  \alpha)  }\rangle
$ and $\vert \psi_{AB}^{(  \alpha^{\prime})  }\rangle $
can contain the same states? Suppose that $\vert \psi_{AB}^{(
\alpha)  }\rangle $ and $\vert \psi_{AB}^{(
\alpha^{\prime})  }\rangle $ are
\begin{equation}
\left\{
\begin{array}
[c]{l}%
\left\vert \psi_{AB}^{\left(  \alpha\right)  }\right\rangle =s_{p+}^{\left(
\alpha\right)  }\left\vert +\phi_{p}\right\rangle \left\vert \eta
_{p+}\right\rangle +s_{p-}^{\left(  \alpha\right)  }\left\vert -\phi
_{p}\right\rangle \left\vert \eta_{p-}\right\rangle ,\\ [6pt]
\left\vert \psi_{AB}^{(  \alpha^{\prime})  }\right\rangle
=s_{p+}^{(  \alpha^{\prime})  }\left\vert +\phi_{p}\right\rangle
\left\vert \eta_{p+}\right\rangle +s_{p-}^{(  \alpha^{\prime})
}\left\vert -\phi_{p}\right\rangle \left\vert \eta_{p-}\right\rangle ,
\end{array}
\right.  \label{eq3.7}%
\end{equation}
with $p\in\{1,2,\cdots,2^{m}\}$. The pure state requirement requires
\begin{equation}
\left\{
\begin{array}
[c]{c}%
s_{p+}^{(  \alpha^{\prime})  }=c_{\alpha^{\prime}\left(  p+\right)
}^{\alpha}s_{p+}^{\left(  \alpha\right)  },\\
s_{p-}^{(  \alpha^{\prime})  }=c_{\alpha^{\prime}\left(  p-\right)
}^{\alpha}s_{p-}^{\left(  \alpha\right)  },
\end{array}
\right.  \label{eq3.8}%
\end{equation}
and
\begin{equation}%
\begin{array}
[c]{l}%
V_{jp+}s_{p+}^{\left(  \alpha\right)  }\left\vert \eta_{p+}\right\rangle
+V_{jp-}s_{ip-}^{\left(  \alpha\right)  }\left\vert \eta_{p-}\right\rangle \\
=d_{\alpha^{\prime}j}^{\alpha}\left(  V_{jp+}s_{p+}^{(  \alpha^{\prime
})  }\left\vert \eta_{p+}\right\rangle +V_{jp-}s_{p-}^{(
\alpha^{\prime})  }\left\vert \eta_{p-}\right\rangle \right)  .
\end{array}
\label{eq3.9}%
\end{equation}
The measurement requirement requires $\vert \eta_{p+}^{(
\alpha)  }\rangle \neq\vert \eta_{p-}^{(  \alpha)
}\rangle $. According to Eq. (\ref{eq3.8}) and Eq. (\ref{eq3.9}), we
have
\begin{equation}
c_{\alpha^{\prime}\left(  p+\right)  }^{\alpha}=c_{\alpha^{\prime}\left(
p-\right)  }^{\alpha}=\frac{1}{d_{\alpha^{\prime}j}^{\alpha}}\equiv
c_{\alpha^{\prime}}^{\alpha}.
\end{equation}
So%
\begin{equation}
\left\vert \psi_{AB}^{(  \alpha^{\prime})  }\right\rangle
=c_{\alpha^{\prime}}^{\alpha}\left\vert \psi_{AB}^{\left(  \alpha\right)
}\right\rangle ,
\end{equation}
that means $\vert \psi_{AB}^{(  \alpha)  }\rangle $ and
$\vert \psi_{AB}^{(  \alpha^{\prime})  }\rangle $ are
the same state. In summary, $\vert \psi_{AB}^{(  \alpha)
}\rangle $ and $\vert \psi_{AB}^{(  \alpha^{\prime})
}\rangle $ cannot contain the same state.

Therefore, for arbitrary $\alpha$ and $\alpha^{\prime}$, we have
\begin{equation}
\left\{
\begin{array}
[c]{l}%
\left\vert \psi_{AB}^{\left(  \alpha\right)  }\right\rangle =s_{q+}^{\left(
\alpha\right)  }\left\vert +\phi_{q}\right\rangle \left\vert \eta
_{q+}\right\rangle +s_{q-}^{\left(  \alpha\right)  }\left\vert -\phi
_{q}\right\rangle \left\vert \eta_{q-}\right\rangle ,\\
\left\vert \psi_{AB}^{(  \alpha^{\prime})  }\right\rangle
=s_{q^{\prime}+}^{(  \alpha^{\prime})  }\left\vert +\phi
_{q^{\prime}}\right\rangle \left\vert \eta_{q^{\prime}+}\right\rangle
+s_{q^{\prime}-}^{(  \alpha^{\prime})  }\left\vert -\phi
_{q^{\prime}}\right\rangle \left\vert \eta_{q^{\prime}-}\right\rangle ,
\end{array}
\right.
\end{equation}
with $q\neq q^{\prime}$ and $q,q^{\prime}=1,2,\cdots,2^{M-1}$. And after
Alice's measurement, Bob's results are
\begin{equation}
\left\{
\begin{array}
[c]{l}%
\tilde{\rho}_{a_{q+}}^{\hat{n}_{k}}=p_{\alpha}\left\vert \omega_{q+}^{k\left(
\alpha\right)  }\right\rangle \left\langle \omega_{q+}^{k\left(
\alpha\right)  }\right\vert ,\\ [6pt]
\tilde{\rho}_{a_{q-}}^{\hat{n}_{k}}=p_{\alpha}\left\vert \omega_{q-}^{k\left(
\alpha\right)  }\right\rangle \left\langle \omega_{q-}^{k\left(
\alpha\right)  }\right\vert .
\end{array}
\right.  \label{eq3.10}%
\end{equation}
Similarly, suppose every Bob's state has a LHS description. Bob can check the
following set of $2^{M+1}$ equations:
\begin{equation}
\left\{
\begin{array}
[c]{l}%
\tilde{\rho}_{a_{1+}}^{\hat{n}_{k}}=%
{\displaystyle\sum\limits_{\xi}}
\wp\left(  a_{1+}|\hat{n}_{k},\xi\right)  \wp_{\xi}\rho_{\xi},\\
\cdots\\
\tilde{\rho}_{a_{2^{M-1}+}}^{\hat{n}_{k}}=%
{\displaystyle\sum\limits_{\xi}}
\wp\left(  a_{2^{M-1}+}|\hat{n}_{k},\xi\right)  \wp_{\xi}\rho_{\xi},\\
\tilde{\rho}_{a_{1-}}^{\hat{n}_{k}}=%
{\displaystyle\sum\limits_{\xi}}
\wp\left(  a_{1-}|\hat{n}_{k},\xi\right)  \wp_{\xi}\rho_{\xi},\\
\cdots\\
\tilde{\rho}_{a_{2^{M-1}-}}^{\hat{n}_{k}}=%
{\displaystyle\sum\limits_{\xi}}
\wp\left(  a_{2^{M-1}-}|\hat{n}_{k},\xi\right)  \wp_{\xi}\rho_{\xi}.
\end{array}
\right.
\end{equation}
According to Eq. (\ref{eq3.10}) , Bob's unnormalized conditional states are
pure. A density matrix of pure state can only be expanded by itself,
therefore, from Eq. (\ref{eq3.10}) we have
\begin{equation}
\left\{
\begin{array}
[c]{l}%
\tilde{\rho}_{a_{1+}}^{\hat{n}_{k}}=\wp_{1}\rho_{1},\\
\cdots\\
\tilde{\rho}_{a_{2^{M-1}+}}^{\hat{n}_{k}}=\wp_{2^{M-1}}\rho_{2^{M-1}},\\
\tilde{\rho}_{a_{1-}}^{\hat{n}_{k}}=\wp_{2^{M-1}+1}\rho_{2^{M-1}+1},\\
\cdots\\
\tilde{\rho}_{a_{2^{M-1}-}}^{\hat{n}_{k}}=\wp_{2^{M}}\rho_{2^{M}}.
\end{array}
\right.  k=1,2
\end{equation}
By summing them up and taking the trace, the left side gives $\mathrm{tr}%
(  \tilde{\rho}^{\hat{n}_{1}}+\tilde{\rho}^{\hat{n}_{2}})
=2\mathrm{tr}(  \rho_{B})  =2$. But the right side gives
$\mathrm{tr}(  \wp_{1}\rho_{1}+\cdots+\wp_{2^{M+1}}\rho_{2^{M+1}})
=\mathrm{tr}(  \rho_{B})  =1$. This leads to the contradiction
\textquotedblleft2=1\textquotedblright

In summary, we discuss the steering paradox with the Bell-like basis
measurement. And it shows that for Bell-like basis measurement, when
$\rho_{AB}$ satisfied both the pure state requirement and the measurement
requirement, we can obtain the contradiction \textquotedblleft%
2=1\textquotedblright. In this case, $\vert \psi_{AB}^{(
\alpha)  }\rangle $ and $\vert \psi_{AB}^{(
\alpha^{\prime})  }\rangle $ cannot contain the same state, and
only contain two items. It is evident that the maximum value range of $\alpha$
is from $1$ to $2^{M-1}$ ($M$ is the number of particles of Alice), that is,
the maximum rank of $\rho_{AB}$ is $2^{M-1}$ for the Bell-like basis measurement.

\section{Nonexistence of contradiction ``2 = 1'' for 2-qubit mixed states}

In this section, we apply the results to the 2-qubit mixed states. We get the
following corollary.

\begin{corollary}
If Alice and Bob share a 2-qubit mixed state, there is no contradiction
\textquotedblleft2 = 1\textquotedblright.
\end{corollary}

From the above analysis, we know that for the Bell-like basis measurement, the
maximum rank of $\rho_{AB}$ is $2^{M-1}$. Suppose Alice and Bob share a
2-qubit entangled state $\rho_{AB}$ and each of them has a qubit, then $M=1$.
If Alice performs the Bell-like basis measurement, the maximum rank of
$\rho_{AB}$ is $1$. That is, for 2-qubit entangled states with the Bell-like
basis measurement, only the $2$-qubit prue entangled states have the
contradiction \textquotedblleft2=1\textquotedblright. So is there a
contradiction \textquotedblleft2=1\textquotedblright\ for 2-qubit mixed
states? Next, we prove that there is no contradiction \textquotedblleft2 =
1\textquotedblright\ for 2-qubit mixed states

\begin{proof}
Suppose Alice and Bob share a 2-qubit mixed state $\rho_{AB}=\sum_{\alpha
}p_{\alpha}\vert \psi_{AB}^{(  \alpha)  }\rangle
\langle \psi_{AB}^{(  \alpha)  }\vert $, in which
\begin{equation}
\left\vert \psi_{AB}^{\left(  \alpha\right)  }\right\rangle =s_{1}^{\left(
\alpha\right)  }\left\vert \phi_{1}\right\rangle \left\vert \eta
_{1}\right\rangle +s_{2}^{\left(  \alpha\right)  }\left\vert \phi
_{2}\right\rangle \left\vert \eta_{2}\right\rangle .\label{eq3.2.1}%
\end{equation}
Here we take $\vert \eta_{1}^{(  \alpha)  }\rangle $ $=$
$\vert \eta_{1}\rangle $ and $\vert \eta_{2}^{(
\alpha)  }\rangle $ $=$ $\vert \eta_{2}\rangle $ to
satisfy the pure state requirement. And the measurement requirement requests
$s_{1}^{(  \alpha)  },s_{2}^{(  \alpha)  }\neq0$, and
$\vert \eta_{1}\rangle \neq\vert \eta_{2}\rangle $.
Alice takes one particle and the other belongs to Bob. Similaily, in the
2-setting steering protocol $\{  \hat{n}_{1},\hat{n}_{2}\}  $,
suppose that Alice' s projective measurements are $\{  \vert
\phi_{i}\rangle \langle \phi_{i}\vert \}  $ and
$\{  \vert \varphi_{j}\rangle \langle \varphi
_{j}\vert \}  $, where $\vert \langle \phi_{i}%
|\varphi_{j}\rangle \vert <1$ and $i,j=1,2$. Then Bob's
unnormalized conditional states are
\begin{equation}
\left\{
\begin{array}
[c]{l}%
\tilde{\rho}_{a_{i}}^{\hat{n}_{1}}=\sum_{\alpha}p_{\alpha}\left\vert
s_{i}^{\left(  \alpha\right)  }\right\vert ^{2}\left\vert \eta_{i}%
\right\rangle \left\langle \eta_{i}\right\vert ,\\ [6pt]
\tilde{\rho}_{a_{j}^{\prime}}^{\hat{n}_{2}}=\sum_{\alpha}p_{\alpha}\left\vert
\chi_{j}^{\left(  \alpha\right)  }\right\rangle \left\langle \chi_{j}^{\left(
\alpha\right)  }\right\vert ,
\end{array}
\right.  \label{eq3.2.2}%
\end{equation}
in which $\vert \chi_{j}^{\left(  \alpha\right)  }\rangle =\sum
_{i}V_{ji}s_{i}^{\left(  \alpha\right)  }\vert \eta_{i}^{\left(
\alpha\right)  }\rangle $, $V_{ji}=\left\langle \varphi_{j}|\phi
_{i}\right\rangle $. In order to get the contradiction, for any $\alpha$ and
$\alpha^{\prime}$, the conditional state of Bob needs to satisfy
\begin{equation}
\left\{
\begin{array}
[c]{l}%
\left\vert \chi_{j}^{\left(  \alpha\right)  }\right\rangle =d_{\alpha^{\prime
}j}^{\alpha}\left\vert \chi_{j}^{(  \alpha^{\prime})
}\right\rangle ,\\ [6pt]
s_{i}^{\left(  \alpha\right)  }=c_{\alpha^{\prime}i}^{\alpha}s_{i}^{\left(
\alpha^{\prime}\right)  }.
\end{array}
\right.  \label{eq3.2.3}%
\end{equation}
According to Eq. (\ref{eq3.2.3}) , we have $c_{\alpha^{\prime}(
i)  }^{\alpha}=c_{\alpha^{\prime}(  i)  }^{\alpha}=
1/{d_{\alpha^{\prime}j}^{\alpha}}$. Then we can obtain $\vert \psi
_{AB}^{(  \alpha)  }\rangle $ and $\vert \psi
_{AB}^{(  \alpha^{\prime})  }\rangle $ are the same state,
similarly. Therefore, there is no contradiction \textquotedblleft%
2=1\textquotedblright\ in 2-qubit mixed states.
\end{proof}

\section{Conclusion}

We have presented a simple EPR\ steering paradox that shows the
incompatibility of the local-hidden-state models with quantum theory for any
$N$-qubit entangled state based on a 2-setting steering protocol. The argument
is valid for any $N$-qubit entangled state, not only $N$-qubit pure entangled
states, but more importantly, $N$-qubit mixed entangled states. We propose a
simple theorem and prove that for any $N$-qubit state satisfying
simultaneously \textquotedblleft the pure state requirement\textquotedblright%
\ and \textquotedblleft the measurement requirement\textquotedblright, then
the contradiction \textquotedblleft2=1\textquotedblright\ can be obtained. In
the example of Bell-like basis measurement, we obtain that the maximum rank of
the $N$-qubit mixed state is $2^{M-1}$ ($M$ is the number of particles of
Alice), and prove that there is no contradiction \textquotedblleft%
2=1\textquotedblright\ in the 2-qubit mixed state. In general, we obtain the
contradiction \textquotedblleft2=1\textquotedblright\ in a more general case.
Furthermore, if one considers the EPR steering scenario in $k$-setting for 
arbitrary $N$-qudit entangled mixed state, then following the similar approach 
one can arrive at a full contradiction, i.e., ``$k= 1$''.

\begin{acknowledgments}
J.L.C. is supported by the National Natural Science Foundation of China (Grants No. 12275136 and 12075001) and the 111 Project of B23045.
H.X.M. is supported by the National Natural Science Foundation of China (Grant No. 11901317). Z.J.L. is supported by the Nankai Zhide Foundation.
\end{acknowledgments}

\newpage

\appendix

\begin{widetext}

   \section{Can Bob have the same state in the same measurement?}

   In order to obtain the contradiction ``2=1'', we analyze whether Bob can have
   the same pure state in the same measurement, and whether Bob can have the same
   pure state in different measurements.
   
   After Alice's measurement, Bob obtains
   \begin{equation}
      \left\{
      \begin{array}
      [c]{l}%
      \tilde{\rho}_{a_{1+}}^{\hat{n}_{1}}=\sum_{\alpha}p_{\alpha}\left\vert
      s_{1+}^{\left(  \alpha\right)  }\right\vert ^{2}\left\vert \eta_{1+}%
      \right\rangle \left\langle \eta_{1+}\right\vert ,\\
      \cdots\\
      \tilde{\rho}_{a_{m+}}^{\hat{n}_{1}}=\sum_{\alpha}p_{\alpha}\left\vert
      s_{m+}^{\left(  \alpha\right)  }\right\vert ^{2}\left\vert \eta_{m+}%
      \right\rangle \left\langle \eta_{m+}\right\vert ,\\
      \cdots\\
      \tilde{\rho}_{a_{n+}}^{\hat{n}_{1}}=\sum_{\alpha}p_{\alpha}\left\vert
      s_{n+}^{\left(  \alpha\right)  }\right\vert ^{2}\left\vert \eta_{n+}%
      \right\rangle \left\langle \eta_{n+}\right\vert ,\\
      \cdots\\
      \tilde{\rho}_{a_{2^{M-1}+}}^{\hat{n}_{1}}=\sum_{\alpha}p_{\alpha}\left\vert
      s_{2^{M-1}+}^{\left(  \alpha\right)  }\right\vert ^{2}\left\vert \eta
      _{2^{M-1}+}\right\rangle \left\langle \eta_{2^{M-1}+}\right\vert ,\\
      \tilde{\rho}_{a_{2^{M-1}-}}^{\hat{n}_{1}}=\sum_{\alpha}p_{\alpha}\left\vert
      s_{2^{M-1}-}^{\left(  \alpha\right)  }\right\vert ^{2}\left\vert \eta
      _{2^{M-1}-}\right\rangle \left\langle \eta_{2^{M-1}-}\right\vert ,\\
      \cdots\\
      \tilde{\rho}_{a_{n-}}^{\hat{n}_{1}}=\sum_{\alpha}p_{\alpha}\left\vert
      s_{n-}^{\left(  \alpha\right)  }\right\vert ^{2}\left\vert \eta_{n-}%
      \right\rangle \left\langle \eta_{n-}\right\vert ,\\
      \cdots\\
      \tilde{\rho}_{a_{m-}}^{\hat{n}_{1}}=\sum_{\alpha}p_{\alpha}\left\vert
      s_{m-}^{\left(  \alpha\right)  }\right\vert ^{2}\left\vert \eta_{m-}%
      \right\rangle \left\langle \eta_{m-}\right\vert ,\\
      \cdots\\
      \tilde{\rho}_{a_{1-}}^{\hat{n}_{1}}=\sum_{\alpha}p_{\alpha}\left\vert
      s_{1-}^{\left(  \alpha\right)  }\right\vert ^{2}\left\vert \eta_{1-}%
      \right\rangle \left\langle \eta_{1-}\right\vert ,\\
      \tilde{\rho}_{a_{1+}^{\prime}}^{\hat{n}_{2}}=\sum_{\alpha}p_{\alpha}\left\vert
      t_{1+}^{\left(  \alpha\right)  }\right\vert ^{2}\left\vert \varepsilon
      _{1+}\right\rangle \left\langle \varepsilon_{1+}\right\vert ,\\
      \cdots\\
      \tilde{\rho}_{a_{m^{\prime}+}^{\prime}}^{\hat{n}_{2}}=\sum_{\alpha}p_{\alpha
      }\left\vert s_{m^{\prime}+}^{\left(  \alpha\right)  }\right\vert
      ^{2}\left\vert \varepsilon_{m^{\prime}+}\right\rangle \left\langle
      \varepsilon_{m^{\prime}+}\right\vert ,\\
      \cdots\\
      \tilde{\rho}_{a_{n^{\prime}+}^{\prime}}^{\hat{n}_{2}}=\sum_{\alpha}p_{\alpha
      }\left\vert s_{n^{\prime}+}^{\left(  \alpha\right)  }\right\vert
      ^{2}\left\vert \varepsilon_{n^{\prime}+}\right\rangle \left\langle
      \varepsilon_{n^{\prime}+}\right\vert ,\\
      \cdots\\
      \tilde{\rho}_{a_{2^{M-1}+}^{\prime}}^{\hat{n}_{2}}=\sum_{\alpha}p_{\alpha
      }\left\vert t_{2^{M-1}+}^{\left(  \alpha\right)  }\right\vert ^{2}\left\vert
      \varepsilon_{2^{M-1}+}\right\rangle \left\langle \varepsilon_{2^{M-1}%
      +}\right\vert ,\\
      \tilde{\rho}_{a_{2^{M-1}-}^{\prime}}^{\hat{n}_{2}}=\sum_{\alpha}p_{\alpha
      }\left\vert t_{2^{M-1}-}^{\left(  \alpha\right)  }\right\vert ^{2}\left\vert
      \varepsilon_{2^{M-1}-}\right\rangle \left\langle \varepsilon_{2^{M-1}%
      -}\right\vert ,\\
      \cdots\\
      \tilde{\rho}_{a_{n^{\prime}-}^{\prime}}^{\hat{n}_{2}}=\sum_{\alpha}p_{\alpha
      }\left\vert s_{n^{\prime}-}^{\left(  \alpha\right)  }\right\vert
      ^{2}\left\vert \varepsilon_{n^{\prime}-}\right\rangle \left\langle
      \varepsilon_{n^{\prime}-}\right\vert ,\\
      \cdots\\
      \tilde{\rho}_{a_{m^{\prime}-}^{\prime}}^{\hat{n}_{2}}=\sum_{\alpha}p_{\alpha
      }\left\vert s_{m^{\prime}-}^{\left(  \alpha\right)  }\right\vert
      ^{2}\left\vert \varepsilon_{m^{\prime}-}\right\rangle \left\langle
      \varepsilon_{m^{\prime}-}\right\vert ,\\
      \cdots\\
      \tilde{\rho}_{a_{1-}^{\prime}}^{\hat{n}_{2}}=\sum_{\alpha}p_{\alpha}\left\vert
      t_{1-}^{\left(  \alpha\right)  }\right\vert ^{2}\left\vert \varepsilon
      _{1-}\right\rangle \left\langle \varepsilon_{1-}\right\vert .
      \end{array}
      \right.  \label{A1}%
      \end{equation}

   In Eq. (\ref{A1}), assuming that $\left\vert \eta_{m+}\right\rangle
   =\left\vert \eta_{n+}\right\rangle $ , i.e., only $2^{M+1}-1$ different pure
   states appear in the quantum result of Eq. (\ref{A1}), so that it is
   sufficient to take $\xi$ from $1$ to $2^{M+1}-1$, namely, one can take the
   ensemble as
   \begin{equation}
   \left\{  \wp_{\xi}\rho_{\xi}\right\}  =\left\{  \wp_{1}\rho_{1},\wp_{2}%
   \rho_{2},\wp_{3}\rho_{3},\cdots,\wp_{2^{M+1}-1}\rho_{2^{M+1}-1}\right\}
   .\label{A2}%
   \end{equation}
   Then Eq. (\ref{A1}) can be written as
   \begin{equation}
   \left\{
   \begin{array}
   [c]{l}%
   \tilde{\rho}_{a_{1+}}^{\hat{n}_{1}}=%
   {\displaystyle\sum\limits_{\xi=1}^{2^{M+1}-1}}
   \wp\left(  a_{1+}|\hat{n}_{1},\xi\right)  \wp_{\xi}\rho_{\xi},\\
   \cdots\\
   \tilde{\rho}_{a_{m+}}^{\hat{n}_{1}}=%
   {\displaystyle\sum\limits_{\xi=1}^{2^{M+1}-1}}
   \wp\left(  a_{m+}|\hat{n}_{1},\xi\right)  \wp_{\xi}\rho_{\xi},\\
   \cdots\\
   \tilde{\rho}_{a_{n+}}^{\hat{n}_{1}}=%
   {\displaystyle\sum\limits_{\xi=1}^{2^{M+1}-1}}
   \wp\left(  a_{n+}|\hat{n}_{1},\xi\right)  \wp_{\xi}\rho_{\xi},\\
   \cdots\\
   \tilde{\rho}_{a_{1-}}^{\hat{n}_{1}}=%
   {\displaystyle\sum\limits_{\xi=1}^{2^{M+1}-1}}
   \wp\left(  a_{1-}|\hat{n}_{1},\xi\right)  \wp_{\xi}\rho_{\xi},\\
   \tilde{\rho}_{a_{1+}^{\prime}}^{\hat{n}_{2}}=%
   {\displaystyle\sum\limits_{\xi=1}^{2^{M+1}-1}}
   \wp\left(  a_{1+}^{\prime}|\hat{n}_{2},\xi\right)  \wp_{\xi}\rho_{\xi},\\
   \cdots\\
   \tilde{\rho}_{a_{1-}^{\prime}}^{\hat{n}_{2}}=%
   {\displaystyle\sum\limits_{\xi=1}^{2^{M+1}-1}}
   \wp\left(  a_{1-}^{\prime}|\hat{n}_{2},\xi\right)  \wp_{\xi}\rho_{\xi}.
   \end{array}
   \right.  \label{A3}%
   \end{equation}
   Since the $2^{M+1}$ states on the left-hand side of Eq. (\ref{A1}) are all
   pure states, a pure state cannot be obtained by a convex sum of other
   different states. Therefore,
   \begin{equation}
   \left.
   \begin{array}
   [c]{l}%
   \tilde{\rho}_{a_{1+}}^{\hat{n}_{1}}=\wp\left(  a_{1+}|\hat{n}_{1},1\right)
   \wp_{1}\rho_{1},\\
   \wp\left(  a_{1+}|\hat{n}_{1},2\right)  =\wp\left(  a_{1+}|\hat{n}%
   _{1},3\right)  =\cdots=\wp\left(  a_{1+}|\hat{n}_{1},2^{M+1}-1\right)  =0.
   \end{array}
   \right.
   \end{equation}
   Similarly, one has
   \begin{equation}
   \left.
   \begin{array}
   [c]{l}%
   \tilde{\rho}_{a_{2+}}^{\hat{n}_{1}}=\wp\left(  a_{2+}|\hat{n}_{1},2\right)
   \wp_{2}\rho_{2},\\
   \wp\left(  a_{2+}|\hat{n}_{1},1\right)  =\wp\left(  a_{2+}|\hat{n}%
   _{1},3\right)  =\cdots=\wp\left(  a_{2+}|\hat{n}_{1},2^{M+1}-1\right)  =0.
   \end{array}
   \right.
   \end{equation}%
   \[
   \cdots
   \]%
   \begin{equation}
   \left.
   \begin{array}
   [c]{l}%
   \tilde{\rho}_{a_{m+}}^{\hat{n}_{1}}=\wp\left(  a_{m+}|\hat{n}_{1},m\right)
   \wp_{m}\rho_{m},\\
   \wp\left(  a_{m+}|\hat{n}_{1},1\right)  =\wp\left(  a_{m+}|\hat{n}%
   _{1},2\right)  =\cdots=\wp\left(  a_{m+}|\hat{n}_{1},2^{M+1}-1\right)  =0.
   \end{array}
   \right.
   \end{equation}%
   \[
   \cdots
   \]%
   \begin{equation}
   \left.
   \begin{array}
   [c]{l}%
   \tilde{\rho}_{a_{n+}}^{\hat{n}_{1}}=\wp\left(  a_{n+}|\hat{n}_{1},m\right)
   \wp_{m}\rho_{m},\\
   \wp\left(  a_{n+}|\hat{n}_{1},1\right)  =\wp\left(  a_{n+}|\hat{n}%
   _{1},2\right)  =\cdots=\wp\left(  a_{n+}|\hat{n}_{1},2^{M+1}-1\right)  =0.
   \end{array}
   \right.
   \end{equation}%
   \[
   \cdots
   \]%
   \begin{equation}
   \left.
   \begin{array}
   [c]{l}%
   \tilde{\rho}_{a_{1-}}^{\hat{n}_{1}}=\wp\left(  a_{1-}|\hat{n}_{1}%
   ,2^{M}-1\right)  \wp_{2^{M}-1}\rho_{2^{M}-1},\\
   \wp\left(  a_{2^{M}}|\hat{n}_{1},1\right)  =\wp\left(  a_{2^{M}}|\hat{n}%
   _{1},2\right)  =\cdots=\wp\left(  a_{2^{M}}|\hat{n}_{1},2^{M+1}-1\right)  =0.
   \end{array}
   \right.
   \end{equation}%
   \begin{equation}
   \left.
   \begin{array}
   [c]{l}%
   \tilde{\rho}_{a_{1+}^{\prime}}^{\hat{n}_{2}}=\wp\left(  a_{1+}^{\prime}%
   |\hat{n}_{2},1\right)  \wp_{2^{M}}\rho_{2^{M}},\\
   \wp\left(  a_{1+}^{\prime}|\hat{n}_{2},2\right)  =\wp\left(  a_{1+}^{\prime
   }|\hat{n}_{2},3\right)  =\cdots=\wp\left(  a_{1+}^{\prime}|\hat{n}_{2}%
   ,2^{M+1}-1\right)  =0.
   \end{array}
   \right.
   \end{equation}%
   \[
   \cdots
   \]%
   \begin{equation}
   \left.
   \begin{array}
   [c]{l}%
   \tilde{\rho}_{a_{1-}^{\prime}}^{\hat{n}_{2}}=\wp\left(  a_{1-}^{\prime}%
   |\hat{n}_{2},2^{M+1}-1\right)  \wp_{2^{M+1}-1}\rho_{2^{M+1}-1},\\
   \wp\left(  a_{1-}^{\prime}|\hat{n}_{2},1\right)  =\wp\left(  a_{1-}^{\prime
   }|\hat{n}_{2},2\right)  =\cdots=\wp\left(  a_{1-}^{\prime}|\hat{n}_{2}%
   ,2^{M+1}-2\right)  =0.
   \end{array}
   \right.
   \end{equation}

   Because $%
   {\sum\limits_{a}}
   \wp(  a|\hat{n},\xi)  =1$, one has
   \begin{equation}
   \left\{
   \begin{array}
   [c]{l}%
   \tilde{\rho}_{a_{1+}}^{\hat{n}_{1}}=\wp_{1}\rho_{1},\\
   \cdots\\
   \tilde{\rho}_{a_{m+}}^{\hat{n}_{1}}=\wp\left(  a_{m+}|\hat{n}_{1},m\right)
   \wp_{m}\rho_{m},\\
   \cdots\\
   \tilde{\rho}_{a_{n+}}^{\hat{n}_{1}}=\wp\left(  a_{n+}|\hat{n}_{1},n\right)
   \wp_{n}\rho_{n},\\
   \cdots\\
   \tilde{\rho}_{a_{1-}}^{\hat{n}_{1}}=\wp_{2^{M}-1}\rho_{2^{M}-1},\\
   \tilde{\rho}_{a_{1+}^{\prime}}^{\hat{n}_{2}}=\wp_{2^{M}}\rho_{2^{M}},\\
   \cdots\\
   \tilde{\rho}_{a_{1-}^{\prime}}^{\hat{n}_{2}}=\wp_{2^{M+1}-1}\rho_{2^{M+1}-1}.
   \end{array}
   \right.  \label{A4}%
   \end{equation}
   where $\wp\left(  a_{m+}|\hat{n}_{1},m\right)  +\wp\left(  a_{n+}|\hat{n}%
   _{1},n\right)  =1$, The sum on the left-hand side of Eq. (\ref{A4}) is
   $2\rho_{B}$, and the sum on the right-hand side is $\rho_{B}$. By summing
   terms in Eq. (\ref{A4}) and taking trace, we arrive at the contradiction
   \textquotedblleft2=1\textquotedblright. So in the same measurement, if Bob
   gets the same pure state, we can get the contradiction \textquotedblleft%
   2=1\textquotedblright.
   
   \section{Can Bob have the same state in different measurements?}
   
   In Eq. (\ref{A1}), assuming that $\left\vert \eta_{m+}\right\rangle
   =\left\vert \varepsilon_{m^{\prime}+}\right\rangle $ , i.e., only $2^{M+1}-1$
   different pure states appear in the quantum result of Eq. (\ref{A1}).
   Similarly, it is sufficient to take $\xi$ from $1$ to $2^{M+1}-1$, one can
   take the ensemble as Eq. (\ref{A2}). Then one has Eq. (\ref{A3}). The
   $2^{M+1}$ states on the left-hand side of Eq. (\ref{A1}) are all pure states.
   The same reasoning can be used to obtain%
   
   \begin{equation}
   \left.
   \begin{array}
   [c]{l}%
   \tilde{\rho}_{a_{1+}}^{\hat{n}_{1}}=\wp\left(  a_{1+}|\hat{n}_{1},1\right)
   \wp_{1}\rho_{1},\\
   \wp\left(  a_{1+}|\hat{n}_{1},2\right)  =\wp\left(  a_{1+}|\hat{n}%
   _{1},3\right)  =\cdots=\wp\left(  a_{1+}|\hat{n}_{1},2^{M+1}-1\right)  =0.
   \end{array}
   \right.
   \end{equation}
   Similarly, one has
   \begin{equation}
   \left.
   \begin{array}
   [c]{l}%
   \tilde{\rho}_{a_{2+}}^{\hat{n}_{1}}=\wp\left(  a_{2+}|\hat{n}_{1},2\right)
   \wp_{2}\rho_{2},\\
   \wp\left(  a_{2+}|\hat{n}_{1},1\right)  =\wp\left(  a_{2+}|\hat{n}%
   _{1},3\right)  =\cdots=\wp\left(  a_{2+}|\hat{n}_{1},2^{M+1}-1\right)  =0.
   \end{array}
   \right.
   \end{equation}%
   \[
   \cdots
   \]%
   \begin{equation}
   \left.
   \begin{array}
   [c]{l}%
   \tilde{\rho}_{a_{m+}}^{\hat{n}_{1}}=\wp\left(  a_{m+}|\hat{n}_{1},m\right)
   \wp_{m}\rho_{m},\\
   \wp\left(  a_{m+}|\hat{n}_{1},1\right)  =\wp\left(  a_{m+}|\hat{n}%
   _{1},2\right)  =\cdots=\wp\left(  a_{m+}|\hat{n}_{1},2^{M+1}-1\right)  =0.
   \end{array}
   \right.
   \end{equation}%
   \[
   \cdots
   \]%
   \begin{equation}
   \left.
   \begin{array}
   [c]{l}%
   \tilde{\rho}_{a_{1-}}^{\hat{n}_{1}}=\wp\left(  a_{1-}|\hat{n}_{1}%
   ,2^{M}\right)  \wp_{2^{M}}\rho_{2^{M}},\\
   \wp\left(  a_{2^{M}}|\hat{n}_{1},1\right)  =\wp\left(  a_{2^{M}}|\hat{n}%
   _{1},2\right)  =\cdots=\wp\left(  a_{2^{M}}|\hat{n}_{1},2^{M+1}-1\right)  =0.
   \end{array}
   \right.
   \end{equation}%
   \begin{equation}
   \left.
   \begin{array}
   [c]{l}%
   \tilde{\rho}_{a_{1+}^{\prime}}^{\hat{n}_{2}}=\wp\left(  a_{1+}^{\prime}%
   |\hat{n}_{2},1\right)  \wp_{2^{M}+1}\rho_{2^{M}+1},\\
   \wp\left(  a_{1+}^{\prime}|\hat{n}_{2},2\right)  =\wp\left(  a_{1+}^{\prime
   }|\hat{n}_{2},3\right)  =\cdots=\wp\left(  a_{1+}^{\prime}|\hat{n}_{2}%
   ,2^{M+1}-1\right)  =0.
   \end{array}
   \right.
   \end{equation}%
   \[
   \cdots
   \]%
   \begin{equation}
   \left.
   \begin{array}
   [c]{l}%
   \tilde{\rho}_{a_{_{m^{\prime}+}}^{\prime}}^{\hat{n}_{2}}=\wp\left(
   a_{_{m^{\prime}+}}^{\prime}|\hat{n}_{2},m^{\prime}\right)  \wp_{2^{M}%
   +m^{\prime}}\rho_{2^{M}+m^{\prime}},\\
   \wp\left(  a_{1}^{\prime}|\hat{n}_{2},1\right)  =\wp\left(  a_{1}^{\prime
   }|\hat{n}_{2},2\right)  =\cdots=\wp\left(  a_{1}^{\prime}|\hat{n}_{2}%
   ,2^{M+1}-1\right)  =0.
   \end{array}
   \right.
   \end{equation}%
   \[
   \cdots
   \]%
   \begin{equation}
   \left.
   \begin{array}
   [c]{l}%
   \tilde{\rho}_{a_{1-}^{\prime}}^{\hat{n}_{2}}=\wp\left(  a_{1-}^{\prime}%
   |\hat{n}_{2},2^{M+1}-1\right)  \wp_{2^{M+1}-1}\rho_{2^{M+1}-1},\\
   \wp\left(  a_{1-}^{\prime}|\hat{n}_{2},1\right)  =\wp\left(  a_{1-}^{\prime
   }|\hat{n}_{2},2\right)  =\cdots=\wp\left(  a_{1-}^{\prime}|\hat{n}_{2}%
   ,2^{M+1}-2\right)  =0.
   \end{array}
   \right.
   \end{equation}

   Because $%
   {\sum\limits_{a}}
   \wp(  a|\hat{n},\xi)  =1$, one has
   \begin{equation}
   \left\{
   \begin{array}
   [c]{l}%
   \tilde{\rho}_{a_{1}}^{\hat{n}_{1}}=\wp_{1}\rho_{1},\\
   \cdots\\
   \tilde{\rho}_{a_{m+}}^{\hat{n}_{1}}=\wp_{m}\rho_{m},\\
   \cdots\\
   \tilde{\rho}_{a_{1-}}^{\hat{n}_{1}}=\wp_{2^{M}}\rho_{2^{M}},\\
   \tilde{\rho}_{a_{1+}^{\prime}}^{\hat{n}_{2}}=\wp_{2^{M}+1}\rho_{2^{M}+1},\\
   \cdots\\
   \tilde{\rho}_{a_{_{m^{\prime}+}}^{\prime}}^{\hat{n}_{2}}=\wp_{2^{M}+m^{\prime
   }}\rho_{2^{M}+m^{\prime}},\\
   \cdots\\
   \tilde{\rho}_{a_{1-}^{\prime}}^{\hat{n}_{2}}=\wp_{2^{M+1}-1}\rho_{2^{M+1}-1}.
   \end{array}
   \right.  \label{A5}%
   \end{equation}
   Here $\wp(  a_{m+}|\hat{n}_{1},m)  =1$, $\wp(  a_{_{m^{\prime
   }+}}^{\prime}|\hat{n}_{2},m^{\prime})  =1$, no more  $\wp(
   a_{m+}|\hat{n}_{1},m)  +\wp(  a_{_{m^{\prime}+}}^{\prime}|\hat
   {n}_{2},m^{\prime})  =1$. The sum on the left-hand side of Eq.
   (\ref{A5}) is $2\rho_{B}$, and the sum on the right-hand side is $\rho_{B}%
   +\wp_{i}\rho_{i}$. By summing terms in Eq. (\ref{A5}) and taking trace, we no
   longer get the contradiction \textquotedblleft2=1\textquotedblright. So in the
   different measurements, if Bob gets the same pure state, we can't get the
   contradiction \textquotedblleft2=1\textquotedblright.
   
   \section{$\left\vert \pm\phi_{i}\right\rangle $ is a set of complete basis of
   Hilbert space}
   
   Here we show that $\vert \pm\phi_{i}\rangle $ ( $i=1,2,\cdots
   ,2^{M-1}$) which in the Bell-like basis measurement is a set of complete basis
   of $2^{M}$-dimensional Hilbert space. For any $k$( $k=1,2$), we have $%
   \sum_{i=1}^{2^{M-1}}
   ( P_{a_{i+}}^{\hat{n}_{k}}+P_{a_{i-}}^{\hat{n}_{k}})  =1$. That
   can be expanded as%
   \begin{align}
   &
   {\displaystyle\sum\limits_{i=1}^{2^{M-1}}}
   \left(  P_{a_{i+}}^{\hat{n}_{k}}+P_{a_{i-}}^{\hat{n}_{k}}\right)  \nonumber\\
   = &
   {\displaystyle\sum\limits_{i=1}^{2^{M-1}}}
   \left(  \cos^{2}\beta_{k}\left\vert +\phi_{i}\right\rangle \left\langle
   +\phi_{i}\right\vert +\cos\beta_{k}\sin\beta_{k}\left\vert +\phi
   _{i}\right\rangle \left\langle -\phi_{i}\right\vert +\sin\beta_{k}\cos
   \beta_{k}\left\vert -\phi_{i}\right\rangle \left\langle +\phi_{i}\right\vert
   +\sin^{2}\beta_{k}\left\vert -\phi_{i}\right\rangle \left\langle -\phi
   _{i}\right\vert \right.  \nonumber\\
   &  \left.  +\sin^{2}\beta_{k}\left\vert +\phi_{i}\right\rangle \left\langle
   +\phi_{i}\right\vert -\sin\beta_{k}\cos\beta_{k}\left\vert +\phi
   _{i}\right\rangle \left\langle -\phi_{i}\right\vert -\cos\beta_{k}\sin
   \beta_{k}\left\vert -\phi_{i}\right\rangle \left\langle +\phi_{i}\right\vert
   +\cos^{2}\beta_{k}\left\vert -\phi_{i}\right\rangle \left\langle -\phi
   _{i}\right\vert \right)  \nonumber\\
   = &
   {\displaystyle\sum\limits_{i=1}^{2^{M-1}}}
   \left(  \left\vert +\phi_{i}\right\rangle \left\langle +\phi_{i}\right\vert
   +\left\vert -\phi_{i}\right\rangle \left\langle -\phi_{i}\right\vert \right)
   \nonumber\\
   = &  {\mathds{1}}%
   \end{align}
   It is obvious that $\vert \pm\phi_{i}\rangle $ is a set of complete
   basis of $2^{M}$-dimensional Hilbert space.%

\end{widetext}

\end{document}